%% file: unidirectionmotion2.tex
\documentclass[aps,twocolumn,groupedaddress, showpacs,prb]{revtex4}
\usepackage{color}
\input{rgb}
\usepackage{graphicx}%

\begin{document}
% You should use BibTeX and apsrev.bst for references
\bibliographystyle{apsrev}

% Use the \preprint command to place your local institutional report
% number on the title page in preprint mode.
% Multiple \preprint commands are allowed.
%\preprint{}

%Title of paper
\title{A traveler-centered intro to kinematics}
% Optional argument for running titles on pages
%\title[]{}

% repeat the \author .. \affiliation  etc. as needed
% \email, \thanks, \homepage, \altaffiliation all apply to the current
% author. Explanatory text should go in the []'s, actual e-mail
% address or url should go in the {}'s for \email and \homepage.
% Please use the appropriate macro for the type of information

% \affiliation command applies to all authors since the last
% \affiliation command. The \affiliation command should follow the
% other informatio
% \affiliation can be followed by \email, \homepage, \thanks as well.
\author{P. Fraundorf}
\affiliation{Physics \& Astronomy/Center for NanoScience, U. Missouri-StL (63121)}
%\affiliation{Physics, Washington University (63110), St. Louis, MO, USA}
\email[]{pfraundorf@umsl.edu}
%\homepage[]{Your web page}
%\thanks{}
%\altaffiliation{}

%Collaboration name if desired (requires use of superscriptaddress
%option in \documentclass). \noaffiliation is required (may also be
%used with the \author command).
%\collaboration can be followed by \email, \homepage, \thanks as well.
%\collaboration{}
%\noaffiliation

\date{\today}

\begin{abstract}

Treating time as a local variable permits robust approaches to kinematics that forego questions of extended-simultaneity, which because of their abstract nature might not be addressed explicitly until a first relativity course and even then without considering the dependence of clock-rates on position in a gravitational field. For example we here use synchrony-free ``traveler kinematic" relations to construct a brief story for beginning students about: (a) time as a local quantity like position that depends on ``which clock", (b) coordinate-acceleration as an approximation to the acceleration felt by a moving traveler, and (c) the geometric origin (hence mass-independence) of gravitational acceleration. The goal is to explicitly rule out global-time for all from the start, so that it can be returned as a local approximation, while tantalizing students interested in the subject with more widely-applicable equations in range of their math background. 

\end{abstract}
% insert suggested PACS numbers in braces on next line
% insert suggested keywords - APS authors don't need to do this
%\keywords{}
%\maketitle must follow title, authors, abstract, \pacs, and \keywords
\maketitle

\tableofcontents

% body of paper here - Use proper section commands+
% References should be done using the \cite, \ref, and \label commands
\section{introduction}
\label{sec:Intro}
%\label{}

Intro-physics students in an engineering-physics class might augment their introduction to unidirectional-kinematics with a technical story that's in harmony with the general trend toward metric-first approaches\cite{Pais82,Taylor01}. This might help nip the implicit Newtonian-assumption of universal time in the bud. It would also introduce: (i) momentum-proportional proper-velocities\cite{SearsBrehme68,Shurcliff96} that can be added vectorially at any speed\cite{Fraundorf2011a}, and (ii) proper-acceleration\cite{Taylor63} in harmony with the modern (equivalence-principle based) understanding of geometric-accelerations i.e. accelerations that arise from choice of a non free-float-frame coordinate-system.

In the process of putting into context the Newtonian-concepts of: {\bf reference-frame}, {\bf elapsed-time} t, {\bf position} x, {\bf velocity} v, {\bf acceleration} a, {\bf constant-acceleration-integral} and {\bf gravitational-acceleration} g, this intro is designed to give students a taste of the more robust technical-concepts highlighted in {\bf bold} below. If these engender critical-discussion (rather than a focus on intution-conflicts), all the better as such concepts might help inspire the empirical-scientist inside students even if they never take another physics course.

The few take-home equations from this introduction that students will likely be asked to master in an intro course will be highlighted in {\color{red}red}. Hence you might simply consider these notes an alternate, but fun, intro to some key relationships that are often just handed to students (explicitly or implicitly) as given. 

\section{time as merely local}

In the first part of the 20th century it was discovered that time is a local variable, linked to each clock's location through a space-time version of Pythagoras' theorem i.e. the local metric equation. Both height in the earth's gravitational field, and clock-motion, affect the rate at which time passes on a given clock. Both of these effects must, for example, be taken into account in the algorithms used by handheld global positioning systems.

The fact that time is local to the clock that's measuring it means that we should probably address the question of extended-simultaneity (i.e. when an event happened from your perspective if you weren't present at the event) only as needed, and with suitable caution. Care is especially needed for events separated by ``space-like'' intervals i.e. for which $\Delta x > c \Delta t$ where $c$ is the space/time constant sometimes called ``lightspeed''.

Recognizing that traveling clocks behave differently also gives us a synchrony-free\cite{Shurcliff96} measure of speed with minimal frame-dependence, namely proper-velocity\cite{SearsBrehme68} $\vec{w} \equiv d\vec{x}/d\tau = \gamma \vec{v}$, which lets us think of momentum as a 3-vector proportional to velocity regardless of speed. Here $\tau$ is the frame-invariant proper-time elapsed on the traveling object's clock, Lorentz-factor $\gamma \equiv dt/d\tau$, and as usual coordinate-velocity $\vec{v} = d\vec{x}/dt$.
    
Recognition of the height-dependence of time as a kinematic (i.e. metric-equation) effect, moreover, allows us to explain the fact that free-falling objects are accelerated by gravity at the same rate regardless of mass. Hence gravity is now seen as a geometric force instead of a proper one, which is only felt from the vantage point of non ``free-float-frame" coordinate-systems like the shell-frame normally inhabited by dwellers on planet earth. 

\section{kinematics teaser}

%\begin{turnpage}
%\begin{figure*}
\begin{figure}
\includegraphics[scale=0.25]{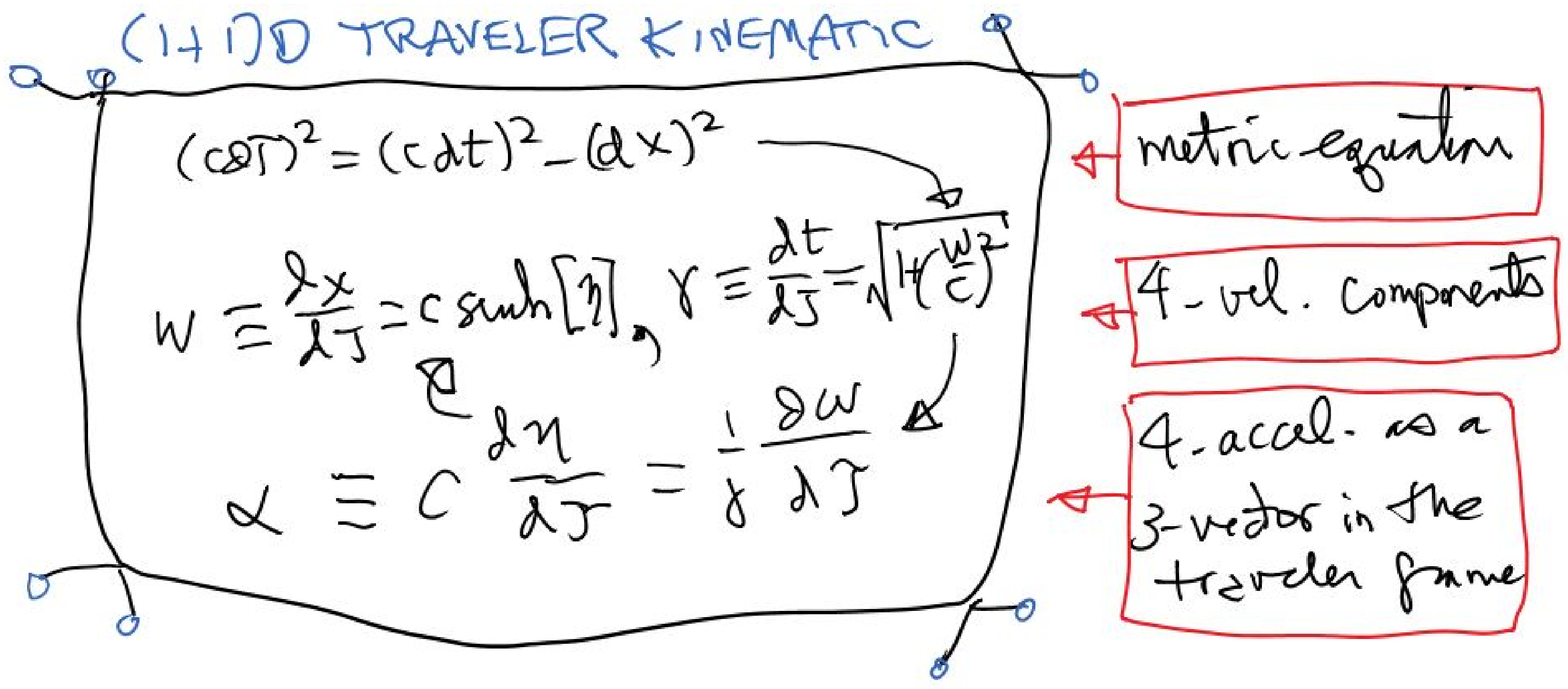}%
\caption{The synchrony-free traveler-kinematic in (1+1)D.}
\label{figA}
\end{figure}
%\end{figure*}
%\end{turnpage}

The world is full of motion, but describing it (that's what kinematics does) requires two perspectives: (i) the perspective of that which is moving e.g. ``the traveler", and (ii) the reference perspective which ain't moving e.g. ``the map". Thus at bare minimum we imagine a map-frame, defined by a coordinate-system of yardsticks say measuring map-position $x$ with synchronized-clocks fixed to those yardsticks measuring map-time $t$, plus a traveler carrying her own clock that measures traveler (or proper) time $\tau$. A definition of extended-simultaneity (i.e. not local to the traveler and her environs), where needed for problems addressed by this approach, is provided by that synchronized array of map-clocks.

We will assume that map-clocks on earth can be synchronized (ignoring the fact that time's rate of passage increases with altitude), but let's initially treat traveler-time $\tau$ as a local quantity that may or may not agree with map-time $t$. The space-time version of Pythagoras' theorem says that in flat space-time, with lightspeed constant $c$, the {\bf Lorentz-factor} or ``speed of map-time" is $\gamma \equiv dt/d\tau = \sqrt{1+(dx/d\tau)^2/c^2}$. This indicates that for many engineering problems on earth (except e.g. for GPS and relativistic-accelerator engineering) we can ignore clock differences, provided we imagine further that {\sc gravity arises} not {\sc from variations in time's passage as a function of height} (i.e. from kinematics) but from a dynamical force that acts on every ounce of an object's being. In that case we can treat time as global, and imagine that accelerations all look the same to observers who are not themselves being accelerated.

Before we take this leap, however, we might spend a paragraph describing kinematics in terms of traveler-centered variables (Fig. \ref{figA}) that allow one to describe motion locally regardless of speed and/or space-time curvature. These variables are frame-invariant {\bf proper-time} $\tau$ on traveler clocks, synchrony-free {\bf proper-velocity} $w \equiv dx/d\tau$ defined in the map-frame, and the frame-invariant {\bf proper-acceleration} $\alpha$ experienced by the traveler, which for unidirectional motion in flat space-time equals $(1/\gamma)dw/d\tau$. Acceleration from the traveler perspective is key, because as Galileo and Newton demonstrated in the 17th century, the causes of motion are intimately connected to this second-derivative of position as a function of time.

%\begin{turnpage}
%\begin{figure*}
\begin{figure}
\includegraphics[scale=0.25]{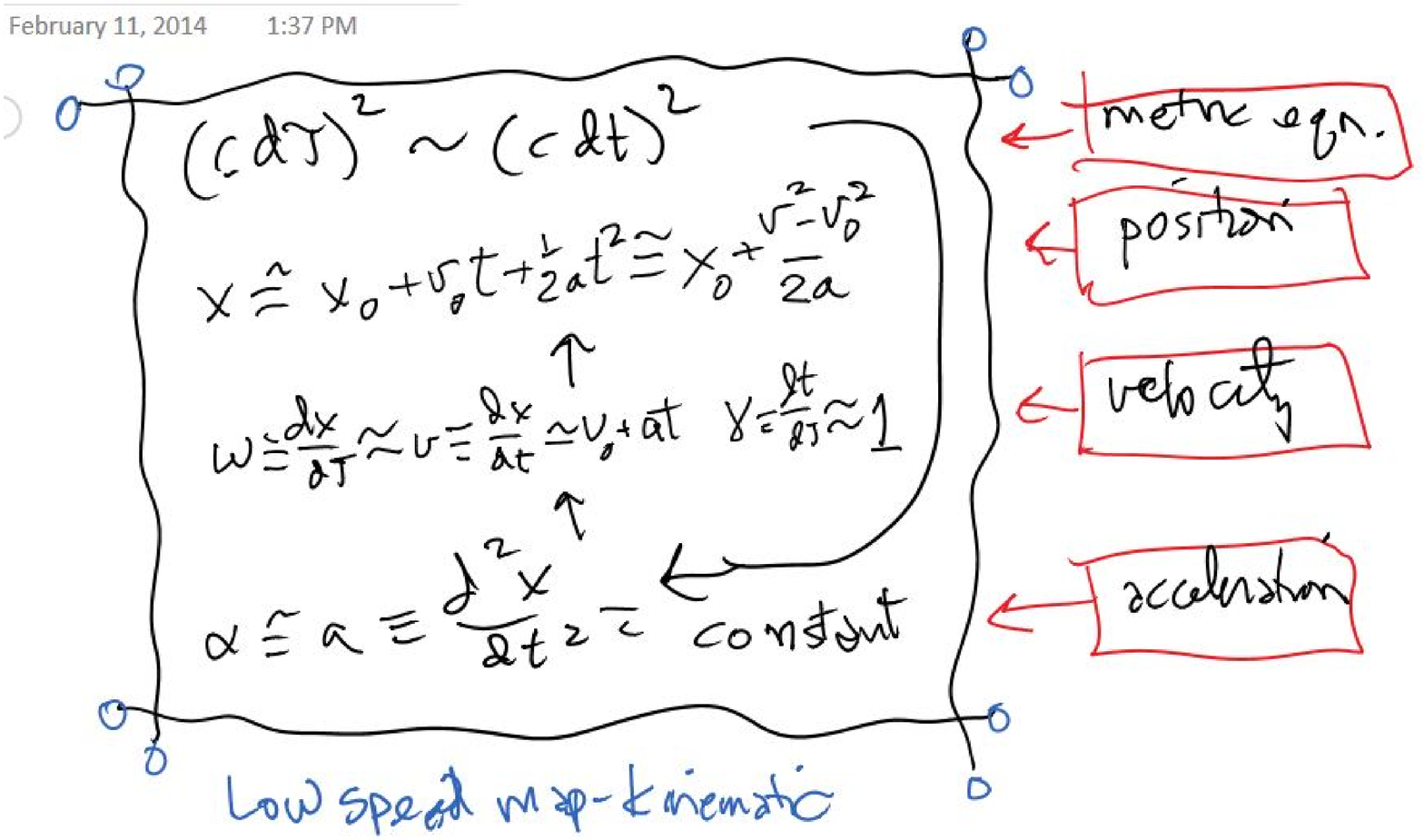}%
\caption{Galileo's approximation to the traveler-kinematic.}
\label{figB}
\end{figure}
%\end{figure*}
%\end{turnpage}

The relationships above allow us to write proper-acceleration as the proper-time derivative of hyperbolic velocity-angle or {\bf rapidity} $\eta$, defined by setting $c \sinh[\eta]$ equal to proper-velocity $w$ in the acceleration direction. These relationships in turn simplify at low speeds (as long as we can also treat space-time as flat) as shown at right below, because one can then approximate the proper-values for velocity and acceleration (Fig. \ref{figB}) with coordinate-values $v \equiv dx/dt$ and $a \equiv dv/dt$.
\begin{equation}
  \begin{array} {rl} 
    {\text{proper-vel.}} & c \sinh[\eta] = \frac{d x}{d \tau}   \\ 
    {\text{proper-accel.}} &  \alpha = c \frac{d \eta}{d \tau} 
  \end{array} 
  \rightarrow
  \begin{array} {rl} 
    v = \frac{d x}{d t}  & {\text{coord.-vel.}} \\ 
    a = \frac{d v}{d t}  & {\text{coord.-accel.}}
  \end{array}  
\end{equation}
As mentioned above, treating space-time as flat requires that our map frame be seen as a free-float-frame (i.e. one experiencing no net forces). Introductory courses therefore concentrate on drawing out uses for the kinematic equations on the right hand side above. We provide the ones on the left, to show that only a bit of added complication will allow one to work in many other situations as well.

\section{constant acceleration for robots}

For a change of pace from most texts, let's discuss the assumptions needed for a {\em computer} to derive the equations of constant acceleration. For equations that work at any speed, we'll also give you some practice treating time as a {\em local} instead of as a global variable i.e. as a value connected to readings on a specific clock.

\subsection{low-speed results}

If we define {\bf coordinate-acceleration} {\color{red}$a \equiv dv/dt$} and {\bf coordinate-velocity} {\color{red}$v \equiv dx/dt$} where $x$ and $t$ are {\bf map-position} and {\bf map-time}, respectively, then holding constant the coordinate-acceleration $a$ (which is not the acceleration felt by our traveler at high speeds) allows one to derive the $ v \ll c$ low-speed constant coordinate-acceleration equations familiar from intro-physics texts for coordinate-velocity $v[t]$ and map-position $x[t]$. Can {\bf you} do it?

The following is what Mathematica needs to pull it off: \\
\\
{\tt FullSimplify[DSolve[\{ \\
\indent v[t] == x'[t], \\
\indent    a == v'[t], \\
\indent    x[0] == xo, \\
\indent    v[0] == vo \\
   \}, \\ 
  \{x[t], v[t]\}, \\
  t \\ 
  ] \\
 ]}.\\

Here we've added the intial ($t=0$) boundary-conditions by defining $x_o$ as initial map-position and $v_o$ as initial coordinate-velocity to eliminate the two constants of integration. Mathematica's output is: \\
\\
{\tt \{ \{ \\
\indent v[t] -> a t + vo, \\
\indent x[t] -> (a t$\wedge$2)/2 + t vo + xo \\
\} \} }.\\

Thus the equations for constant coordinate-acceleration in one direction may be written:
\begin{equation}{\color{red}v = v_o + a t \, }\end{equation}
\begin{equation}{\color{red}x = x_o + v_o  t + \frac12 a t^2 = x_o + \frac{v^2-v_o^2}{2 a}},\end{equation} where as usual $\Delta f \equiv f_{\text{final}}-f_{\text{initial}}$ for any time-varying quantity $f$. Here the first equation tells us how coordinate-velocity $v$ changes with elapsed map-time $t$, while the second tells us how map-position $x$ changes with map-time $t$ as well as with state-of-motion (the work-energy equation) since work is $W \simeq m a \Delta x$ and kinetic energy is $K \simeq \frac12 m v^2$.

In terms of increments instead of differentials for constant unidirectional acceleration, we can therefore also write: $a = \Delta v / \Delta t = \frac12 \Delta[ v^2 ]/ \Delta x$.

\subsection{any-speed results}

%\begin{turnpage}
%\begin{figure*}
\begin{figure}
\includegraphics[scale=0.25]{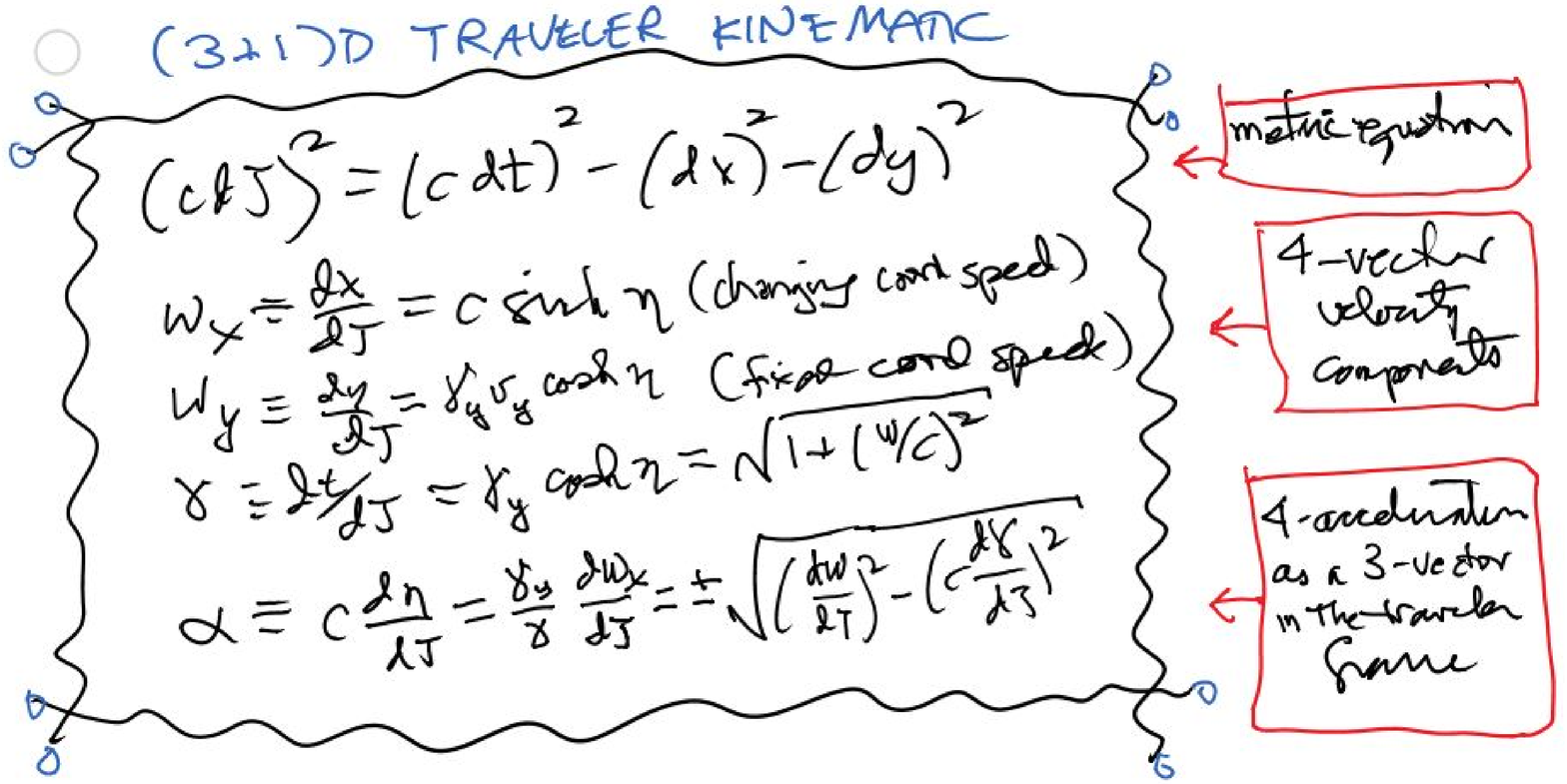}%
\caption{The synchrony-free traveler-kinematic in (3+1)D.}
\label{figC}
\end{figure}
%\end{figure*}
%\end{turnpage}

For equations that work at any speed, we begin by treating the ``proper" time $\tau$ on the clocks of a traveler as a local-variable, whose value we'd like to figure out relative to the local value of the traveler's position $x$ and time $t$ on the yardsticks and synchronized clocks of a reference map-frame. The space-time Pythagorean theorem or ``metric-equation" for flat space-time, namely $(c \delta \tau)^2 = (c \delta t)^2 - (\delta x^2)$ with ``lightspeed" constant $c$, allows us to define minimally frame-variant ``proper" values for the velocity and acceleration, as well as for the time, {\em experienced by our traveler}.

In particular {\bf proper-velocity} $w$ (map-distance $x$ per unit {\bf proper-time} $\tau$) is just $w \equiv dx/d\tau \equiv c \sinh [ \eta  ]$, where $\eta$ is referred to as hyperbolic velocity-angle or {\bf rapidity}. The frame-invariant {\bf proper-acceleration} $\alpha$ felt by a traveler equals this ``length-contracted" proper-velocity derivative  $(1/\gamma)d^2x/d\tau^2 = c d \eta /d t$ i.e. constant $c$ times the traveler-time $\tau$ derivative of rapidity $\eta$. Holding $\alpha$ fixed thus allows one to derive constant proper-acceleration equations that work at any speed for map-position $x[\tau]$ and proper-velocity $w[\tau]$.

In the above discussion we are using proper-time $\tau$ {\em local to the traveler's clocks} as the independent variable, so as to avoid thinking of time as a global variable. Note that unlike coordinate-velocity $v \equiv d x / d t$, proper-velocity $w \equiv d x / d \tau$ always equals momentum per unit mass and has no upper limit. Can {\bf you} figure how map-position $x$ and proper-velocity $w = c \sinh[\eta]$ depend on traveler-time $\tau$, given this information?

The following is what Mathematica needs to pull it off: \\
\\
{\tt FullSimplify[DSolve[\{ \\
\indent   c Sinh[eta[tau]] == x'[tau],\\
\indent   alpha == c eta'[tau], \\
\indent   x[0] == xo, \\
\indent   eta[0] == etao \\
  \},  \\
 \{x[tau], eta[tau]\}, \\ 
 \{tau\} \\
 ] \\
]}. \\

As before we specify two initial ($\tau = 0$) conditions, in this case for initial map-position $x_o$ and initial rapidity or hyperbolic velocity-angle $\eta_o$. The result is: \\
\\
{\tt \{ \{ \\
\indent   eta[tau] -> etao + (alpha tau)/c, \\
\indent   x[tau] -> (alpha xo + c$\wedge$2 (-Cosh[etao] + \\
\indent \indent Cosh[etao + (alpha tau)/c]))/alpha \\
\} \} }.\\

{\em Note that we also get these bonus relationships:} Traveler-speed on the map can be expressed in several ways, including: {\bf Lorentz-factor} $\gamma \equiv dt/d\tau = \cosh[\eta] = \sqrt{1+(w/c)^2} = 1/\sqrt{1-(v/c)^2}$, where coordinate-velocity $v \equiv w/\gamma = c \tanh[\eta]$. For incremental changes when proper-acceleration is constant and all motion is along that direction we can also write $\alpha = \Delta w /\Delta t = c \Delta \eta/\Delta \tau = c^2 \Delta \gamma/\Delta x = \gamma^3 a$. 

All of the foregoing {\bf assertions are local to the traveler's position in the map-frame} of yardsticks and synchronized clocks. If we use those synchronized clocks to define simulaneity between separated events, the above also tells us about {\bf traveler motion from the perspective of stationary observers anywhere on the map}. Hence these equations are spectacular for exploring constant-acceleration round-trips between stars.

Thus the equations, analogous to the Newtonian ones, for unidirectional constant proper-acceleration at any speed might be written:
\begin{equation}w = c \sinh \left[ \frac{\alpha \tau}{c} + \eta_o \right] = w_o + \alpha \int_0^\tau \gamma[\tau'] d\tau' = w_o + \alpha \Delta t ,\end{equation}
\begin{equation}
x = x_o + \frac{c^2}{\alpha} \left( \cosh \left[ \frac{\alpha \tau}{c} + \eta_o \right] - \cosh \left[ \eta_o \right] \right)  = x_o + \frac{c^2}{\alpha} \Delta \gamma ,
\end{equation}
where again $\Delta f \equiv f_{\text{final}} - f_{\text{initial}}$, and $\eta_o \equiv \sinh^{-1} [w_o/c]$. Here the first equation tells us how proper-velocity $w$ changes with elapsed traveler-time $\tau$ and map-time $\Delta t$, while the second tells us how map-position $x$ changes with traveler-time $\tau$ as well as with state-of-motion (the work-energy equation) since work is $m \alpha \Delta x$ and change-in kinetic energy is $\Delta K = m c^2 \Delta \gamma$ given that $K = (\gamma-1)mc^2$. These results generalize nicely to the (3+1)D case (Fig. \ref{figC}), although more care must be taken than in the Galilean approximation since 3-vector magnitudes show more dependence on observer frame.

At low speeds ($ w \ll c$) of course, map and traveler clock times go at the same rate i.e. $dt \simeq d\tau$, the velocity-parameters (coordinate, proper and angle) are essentially the same i.e. $v \simeq w \simeq c \eta$, coordinate and proper acceleration are about equal i.e. $ a \simeq \alpha$, and the any-speed equations reduce to the low-speed ones discussed above.

\section{cool any-speed applications}

Invariant proper-time $\tau $ is already finding its way into intro-physics
and special-relativity books, e.g. to recognize that the frame a clock
resides in is special when the topic of {\em time elapsed on that clock} comes up.
Proper-velocity $w$ and proper-acceleration $\alpha $ for a traveler are less 
consistently mentioned, but also have uses that may be of interest to introductory
physics teachers. In this section we discuss a few of the possibilities.

\subsection{proper-velocity at home}

Although $v\ll c$, $w\ll c$ and $K\ll mc^{2}$ are all natural inequalities
that define the sub-relativistic regime, where for example $\delta
t\simeq \delta t$, coordinate-velocity $v$ inequalities are not useful
compared to the proper-velocity and kinetic-energy inequalities $w\gg c$ and 
$K\gg mc^{2}$ for defining a supra-relativistic regime, where for example
$K\simeq pc$. Moreover, a proper-velocity of just one lightyear per
traveler year i.e. $w\simeq c$ is a natural dividing point between those
two limiting cases.

Relativistic-particle land-speed records will also be more interesting in
lightyears per traveler year ($w$ in ly/ty), than in lightyears per map year
($v$ in ly/y). For example, a 45 GeV electron accelerated in 1989 by the
Large Electron-Positron Collider (LEP) at Cern would have had a
coordinate-speed $v$ of only about sixty four trillionths shy of 1 ly/y. On
the other hand, its proper-speed would have been around 88,000 ly/ty, a
number much more useful for comparisons from one run to the next.

The story of relative proper-velocities may be even more interesting. 
The elegant symmetry of those low-speed relative coordinate-velocity
3-vector equations ($\vec{v}_{AC}=\vec{v}_{AB}+\vec{v}_{BC}$) is not 
preserved at high speeds, but for proper-velocities it is. In the
unidirectional case, we have the lovely result that Lorentz-factors multiply
while coordinate-factors add i.e. $w_{AC}=\gamma _{AB}\gamma
_{BC}(v_{AB}+v_{BC})$. Voters enjoying land-speed records in ly/ty might
really enjoy hearing about the collider advantage. Accelerating two 45 GeV
electrons and colliding them takes the proper land-speed for a single
electron of 88,000 ly/ty up to a relative collider speed of $w_{AC}\simeq
88,000^{2}(1+1)\sim 1.55\times 10^{10}$ ly/ty. Quite a bargain over one
accelerator, for something like twice the cost.

In the multi-directional case, moreover, a 3-vector equation very similar to
the low-speed velocity equation can be written, namely $\vec{w}_{AC}=(\vec{w}%
_{AB})_{C}+\vec{w}_{BC}$. The only complication is that frame C's view of
the out-of-frame proper-velocity $(\vec{w}_{AB})_{C}$ is in the 
direction of $\vec{w}_{AB}\equiv \gamma _{AB}\vec{v}_{AB}$ but must be
rescaled in magnitude\cite{Fraundorf2011a} before the addition will work.

Another fun, but less practical topic, is that of relativistic traffic
safety which includes games in Mr. Tompkins style universes\cite{Gamow45} where e.g.
lightspeed is 55 mph. In such a universe, would interstate highways still
need speed-limit signs? The answer is yes, and it would moreover be a proper
and not a coordinate velocity limit. 

To see this, simply consider (one at a time) which velocity-measure best reflects 
maximum possible collision-damage in terms of vehicle (i) momentum and (ii) kinetic 
energy, and which measure best reflects minimum chance for collision-avoidance 
in terms of (iii) driver and (iv) pedestrian reaction-time. At all speeds, both 
vehicle momentum and kinetic energy scale nicely with proper-velocity while their 
dependence on coordinate-velocity goes through the roof as $v \rightarrow c$.
Similarly driver reaction-time decreases, as does pedestrian reaction-time after 
the warning photon arrives, in a complementary way with increasing proper-velocity 
but not with coordinate-velocity\cite{Fraundorf2011a}. 

Thus in our $c = 55$ mph universe, limiting travelers to speeds of less than 55 
map-miles {\em per traveler-hour} makes more sense than limiting them to less than 
55 or even 54.991 map-miles {\em per map-hour}. Not only would raising the limit
to 60 mph remain a viable option, but as an added bonus the speedometer-reading for 
proper-velocity divided into destination distance directly answers the question 
that kids in the back seat are asking i.e. ``How long (to me) before we get there?"

\subsection{proper-acceleration at work}

Given that acceleration is not always discussed much in 
special-relativity texts\cite{French68}, one might imagine that 
equations for any-speed acceleration are irrelevant to everyday life. 
On the contrary thanks to proper-acceleration's frame-invariance
and general relativity's equivalence-principle, which allows 
Newton's laws to work locally in accelerated (non-free-float) 
frames with help from non-proper geometric (affine-connection) 
forces that act on every ounce of an object's being, proper-acceleration 
allows one to explain the difference between gravity 
and most other intro-physics forces\cite{Fraundorf2011a} from the 
first day of class.

We discuss this in more detail in the curved space-time sections 
below, where the radar-time definition of extended-simultaneity 
will even allow students to visualize the spacetime-curvature that 
results from everyday accelerations in flat space-time. 
In this subsection, we instead address one potentially-practical 
application of the constant proper-acceleration equations derived by a 
computer in the section above, namely interstellar round-trips 
under constant acceleration.

A simple model for ``one-gee" round-trips might be 
someone doing jumping-jacks. Technically of course, on earth 
at least, these take place in a non-free-float frame in which 
launch is accomplished with help of a proper-acceleration while 
the return trip is accomplished with help of a geometric-force 
that acts on every ounce of one's being.

For interstellar roundtrips, let's imagine (even if 
impractical) that we have a ship capable of ``one-gee" 
proper-acceleration either forward or back for an 
extended time. This is a limiting case, as we are adapted 
to survive proper-accelerations this large (but not 
larger) for extended periods on earth.

The good news, if you simply put in the numbers, is that 
how far one can go in a given amount of elapsed 
traveler-time exceeds the distance one can go in a 
Galilean world (without a finite value for space-time constant $c$) 
by a ridiculous amount. In other words, relativity opens up 
rather than closes down possibilities for interstellar travel 
in terms of time-elapsed on traveler-clocks\cite{Lagoute95}. 
It's the couch-potatoes at home that 
relativity hurts, not the travelers themselves.

For instance, a 57-year 1-gee roundtrip using the low-speed 
(non-relativistic) equations for constant coordinate-acceleration 
above would at most allow one to go about 200 lightyears and back. The 
same 57-year trip using the relativistic equations for constant 
proper-acceleration would take you all the way to Andromeda 
galaxy 2 million lightyears away and back.

The bad news is that carrying on-board fuel (even one-way) 
on these trips will make trips just to nearby stars difficult, 
and the thrust-profile for constant proper-acceleration very 
heavily front-loaded\cite{Fraundorf2011a}. 
Extended times at 1-gee acceleration of course might make collisions
with dust particles (or even hydrogen atoms) at ambient 
speeds a non-trivial problem as well.

\section{curving space-time}

Our analysis in the first section, of time as local to the clocks used to measure it, was in part to distance ourselves as much as possible from a discussion of extended simultaneity. In this section, for dealing with accelerated travelers we choose a radar-time model for extended simultaneity\cite{DolbyGull01} (instead of the tangent free-float-frame model) in order to show students how proper-acceleration curves space-time for the traveler all by itself. This model in hand, the door my open a bit wider to experimentation by interested students with simple gravitational metrics in the spirit of Taylor and Wheeler's ``Exploring Black Holes" text\cite{Taylor01}, whose pre-publication draft-title was ``Scouting Black Holes: Exploring General Relativity with Calculus" likely in part to inspire a closer look by intro-physics students.

In this note, we don't have the opportunity to develop the equations to treat curved space in detail. Instead, therefore, we focus on visualizations and on a few bottom-line relationships that might pique a student's interest.

\subsection{acceleration-related curvature}

%\begin{turnpage}
%\begin{figure*}
\begin{figure}
\includegraphics[scale=0.68]{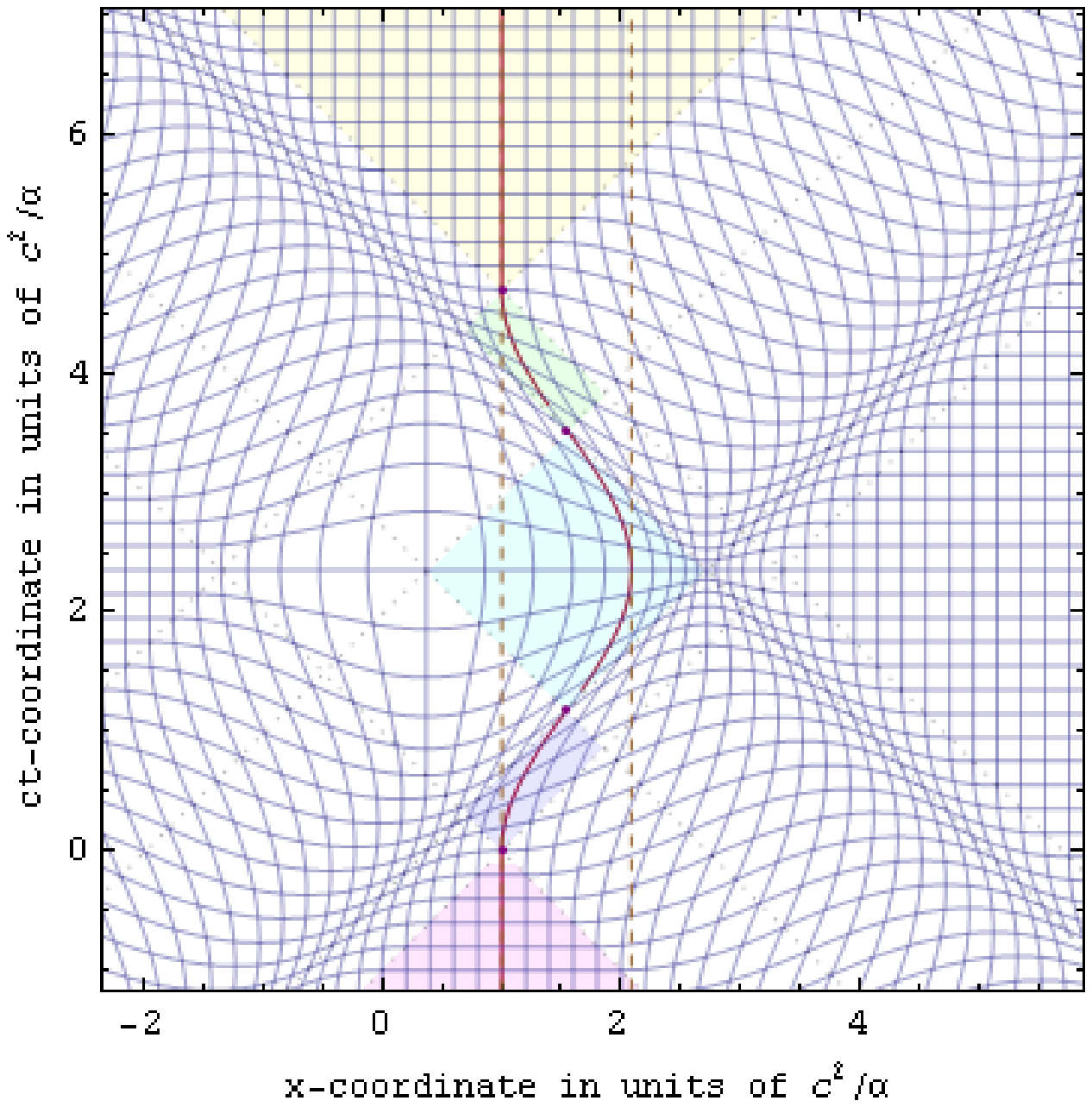}%
\caption{In special-relativity with radar-time simultaneity\cite{DolbyGull01}, acceleration curves flat spacetime.}
\label{fig1}
\end{figure}
%\end{figure*}
%\end{turnpage}

\label{sec:accelerationCurvature}

The map-frame $c t$ versus $x$ radar-time simultaneity plot in Fig. \ref{fig1} shows how acceleration, in this case of a 1-gee proper-acceleration round-trip lasting 4 traveler-years, distorts distances (blue vertical mesh-lines) and simultaneity (blue horizontal mesh-lines) experienced by that accelerated observer.  For objects that are extended along the line of their acceleration, these distortions in space and time will occur even across an accelerated-object's own length.

For example, in addition to the metric-equation's motion-related {\bf time-dilation} in which:
\begin{equation}
\delta \tau_{\text{traveler}} = \delta t_{\text{map}} \sqrt{1-\left(\frac{v}{c}\right)^2},
\label{movingdilation}
\end{equation}
for accelerated objects of length $L$ in the direction of proper-acceleration $\alpha$, one finds an acceleration-related time-dilation of the form:
\begin{equation}
\delta \tau_{\text{trailing}} \simeq \delta \tau_{\text{leading}} \sqrt{1 - \frac{2 \alpha L}{c^2}}.
\label{acceldilation}
\end{equation}
Here the leading-edge of the object is in the direction of the acceleration $\alpha$, not necessarily in the direction of travel.  

\begin{figure}
\includegraphics[scale=0.68]{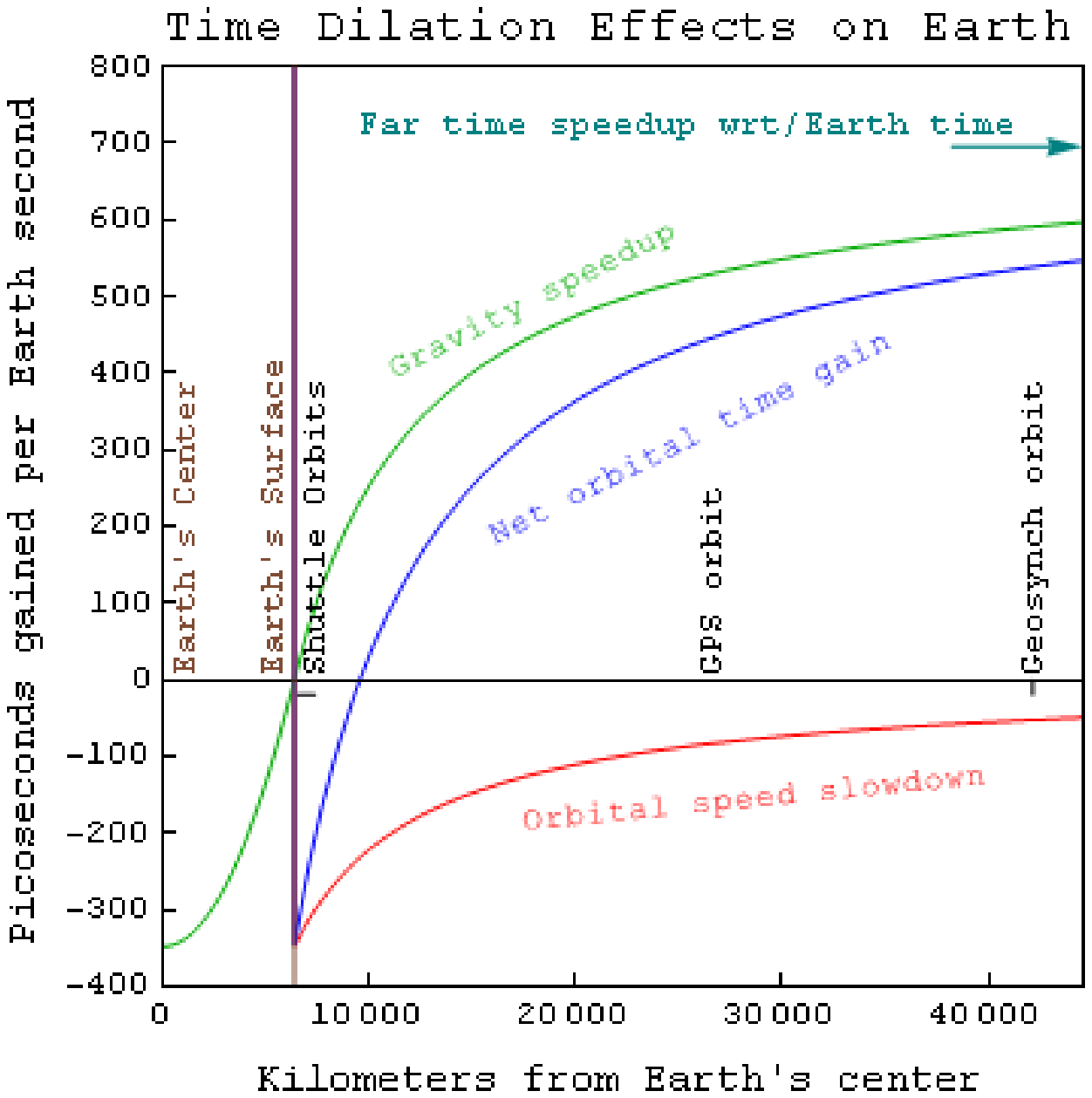}%
\caption{Time-dilation effects in satellite global-positioning.}
\label{fig2}
\end{figure}

For the 1-gee proper-acceleration of a standing human in the vertical direction, this {\bf differential-aging} between head and foot becomes $dt_{\text{foot}}/dt_{\text{head}} \simeq 1 - 2\times 10^{-16}$.  This means that if you stand up (or sit tall) for a sizeable fraction of your lifetime, {\bf your head may be a few-hundred nanoseconds older than your feet}.  This is a small effect for humans, but as discussed below (and illustrated in Fig. \ref{fig2}) it's quite significant for global-positioning satellites for which nano-second timing-errors give rise to macroscopic errors in position.

\subsection{gravity's acceleration}
\label{sec:gees}

%\begin{turnpage}
%\begin{figure*}
%\end{figure*}
%\end{turnpage}

Einstein's general-relativity shows how a gravitational-acceleration that is the same for all masses can be seen to result from a mass-related static-distortion of space-time. This can be described most simply with a modified metric-equation of the form:
\begin{equation}
(c \delta \tau_{\text{radius-r}})^2 = (1-\frac{r_o}{r})(c \delta t_{\text{far}})^2 - \frac{(\delta x_{\text{far}})^2}{(1-\frac{r_o}{r})}. 
\label{gravmetric}
\end{equation}
On earth's surface the metric equation doesn't change by much since $r_o/r \simeq 1.39117\times 10^{-9}$, given that {\bf event-horizon radius} $r_o = 2 G M /c^2$, where $G$ is the universal gravitation constant and $M$ the earth's mass.

Nonetheless, this modified-metric gives rise to {\bf gravity's geometric-acceleration} of $G M /r^2$ that at earth's surface becomes {\color{red}$g \simeq 9.8$ meters per second-squared}, which must be countered by an upward proper-force of $m g$ (as shown later in this course) to keep a {\bf shell-frame} observer's radius fixed in the neighborhood of a planet. That's because shell-frames (of fixed radius) are not free-float-frames. Around gravitational objects, free-float-frames are sometimes called {\bf rain-frames} instead.

The space-time curvature associated with gravity's geometric-acceleration also distorts space and time. One result of this is the gravitational time-dilation of global-positioning-system (GPS) satellites, as well as of your head, relative to your boots on the ground.

As with the previous two expressions for differential-aging, this dilation is also linked to an expression involving $\sqrt{1-2×\text{energy}/mc^2}$, namely:
\begin{equation}
\delta \tau_{\text{radius-r}} \simeq \delta t_{\text{far}} \sqrt{1 - \frac{2 G M}{r c^2}},
\label{grav dilation}
\end{equation} 
where potential-energy per unit mass at radius $r$ (also to be shown later in the course) is $G M /r$. This further means that if clocks at the earth's center and surface began ticking together on the day when earth's formation from the solar-nebula was complete, since then {\bf time-elapsed at earth's center is about a year less than on the surface}. Such differential-aging effects are even more severe with extremely dense objects, like {\bf neutron stars} and the event-horizons of {\bf black holes}.

\section{discussion}
\label{sec:discussion}

We've not provided all the steps needed to arrive at the standard but ``superficially treated'' conclusions described here, both to save space and to leave something for curious students to explore. Many of them follow simply from the corresponding metric. More importantly, the results can often be put to use with only the math-tools required for an introductory course!

One caution that may bear repeating is that like time, simultaneity depends on one's choice of reference frame. The metric-equation variables discussed here (including those for speed and acceleration) implicitly assume {\em simultaneity defined from the vantage point of the map-frame alone}. With that caveat, students inclined to wade through some extra math might be tempted to explore some anyspeed-kinematics as well.

Concerning other applications for the traveler kinematic, four-vectors are of course written in the traveler-kinematic i.e. in terms of derivatives with respect to proper-time. Moreover the {\em free-particle} Lagrangian in curved space-time, when parameterized in terms of proper-time, is simply $-m c^2$. This yields the most elegant and comprehensive prediction of free-particle motion yet: In the absence of proper-acceleration, objects move so as to maximize aging\cite{Taylor2003,Moore2004} i.e. elapsed proper-time. 

The metric-equation strategy used above for relating proper-acceleration to proper-velocity derivatives also works in curved space-times. The problem is that the relation between proper-acceleration and proper-velocity derivatives in curved-space time generally involves a covariant-derivative with connection-coefficients. 

The multi-component tensors required there are beyond the interest of most undergraduates, although a closer look at everyday curvatures (like the gravity around a planet) may be of interest to some. Questions to explore, for instance, might include: (a) how satellite orbits can be predicted by choosing the path of maximal aging, or (b) how the Schwarschild-metric yields within one expression the equations for both shell-frame gravity and accelerated-frame centrifugal-force.

\begin{acknowledgments}
I would like to thank Roger Hill, Bill Shurcliff, and Edwin Taylor 
for their enthusiasm about new ways to look at old things.
\end{acknowledgments}

% If you have acknowledgments, this puts in the proper section head.
%\begin{acknowledgments}
%I would like to thank the late E. T. Jaynes for many interesting papers and 
%discussions over the past half century, and my colleages at the 
%University of Missouri at St. Louis, at Washington University, at Monsanto, 
%at Boeing, and at MEMC Electronic Materials for the opportunity to see 
%these ideas developed and applied in in interdisciplinary environment.
%\end{acknowledgments}

% Create the reference section using BibTeX:

%\begin{references}
\bibliography{ifzx2}

\end{document}

%% file: rgb.tex
\definecolor{AliceBlue}{rgb}{0.94,0.97,1.00}
\definecolor{AntiqueWhite1}{rgb}{1.00,0.94,0.86}
\definecolor{AntiqueWhite2}{rgb}{0.93,0.87,0.80}
\definecolor{AntiqueWhite3}{rgb}{0.80,0.75,0.69}
\definecolor{AntiqueWhite4}{rgb}{0.55,0.51,0.47}
\definecolor{AntiqueWhite}{rgb}{0.98,0.92,0.84}
\definecolor{BlanchedAlmond}{rgb}{1.00,0.92,0.80}
\definecolor{BlueViolet}{rgb}{0.54,0.17,0.89}
\definecolor{CadetBlue1}{rgb}{0.60,0.96,1.00}
\definecolor{CadetBlue2}{rgb}{0.56,0.90,0.93}
\definecolor{CadetBlue3}{rgb}{0.48,0.77,0.80}
\definecolor{CadetBlue4}{rgb}{0.33,0.53,0.55}
\definecolor{CadetBlue}{rgb}{0.37,0.62,0.63}
\definecolor{CornflowerBlue}{rgb}{0.39,0.58,0.93}
\definecolor{DarkBlue}{rgb}{0.00,0.00,0.55}
\definecolor{DarkCyan}{rgb}{0.00,0.55,0.55}
\definecolor{DarkGoldenrod1}{rgb}{1.00,0.73,0.06}
\definecolor{DarkGoldenrod2}{rgb}{0.93,0.68,0.05}
\definecolor{DarkGoldenrod3}{rgb}{0.80,0.58,0.05}
\definecolor{DarkGoldenrod4}{rgb}{0.55,0.40,0.03}
\definecolor{DarkGoldenrod}{rgb}{0.72,0.53,0.04}
\definecolor{DarkGray}{rgb}{0.66,0.66,0.66}
\definecolor{DarkGreen}{rgb}{0.00,0.39,0.00}
\definecolor{DarkGrey}{rgb}{0.66,0.66,0.66}
\definecolor{DarkKhaki}{rgb}{0.74,0.72,0.42}
\definecolor{DarkMagenta}{rgb}{0.55,0.00,0.55}
\definecolor{DarkOliveGreen1}{rgb}{0.79,1.00,0.44}
\definecolor{DarkOliveGreen2}{rgb}{0.74,0.93,0.41}
\definecolor{DarkOliveGreen3}{rgb}{0.64,0.80,0.35}
\definecolor{DarkOliveGreen4}{rgb}{0.43,0.55,0.24}
\definecolor{DarkOliveGreen}{rgb}{0.33,0.42,0.18}
\definecolor{DarkOrange1}{rgb}{1.00,0.50,0.00}
\definecolor{DarkOrange2}{rgb}{0.93,0.46,0.00}
\definecolor{DarkOrange3}{rgb}{0.80,0.40,0.00}
\definecolor{DarkOrange4}{rgb}{0.55,0.27,0.00}
\definecolor{DarkOrange}{rgb}{1.00,0.55,0.00}
\definecolor{DarkOrchid1}{rgb}{0.75,0.24,1.00}
\definecolor{DarkOrchid2}{rgb}{0.70,0.23,0.93}
\definecolor{DarkOrchid3}{rgb}{0.60,0.20,0.80}
\definecolor{DarkOrchid4}{rgb}{0.41,0.13,0.55}
\definecolor{DarkOrchid}{rgb}{0.60,0.20,0.80}
\definecolor{DarkRed}{rgb}{0.55,0.00,0.00}
\definecolor{DarkSalmon}{rgb}{0.91,0.59,0.48}
\definecolor{DarkSeaGreen1}{rgb}{0.76,1.00,0.76}
\definecolor{DarkSeaGreen2}{rgb}{0.71,0.93,0.71}
\definecolor{DarkSeaGreen3}{rgb}{0.61,0.80,0.61}
\definecolor{DarkSeaGreen4}{rgb}{0.41,0.55,0.41}
\definecolor{DarkSeaGreen}{rgb}{0.56,0.74,0.56}
\definecolor{DarkSlateBlue}{rgb}{0.28,0.24,0.55}
\definecolor{DarkSlateGray1}{rgb}{0.59,1.00,1.00}
\definecolor{DarkSlateGray2}{rgb}{0.55,0.93,0.93}
\definecolor{DarkSlateGray3}{rgb}{0.47,0.80,0.80}
\definecolor{DarkSlateGray4}{rgb}{0.32,0.55,0.55}
\definecolor{DarkSlateGray}{rgb}{0.18,0.31,0.31}
\definecolor{DarkSlateGrey}{rgb}{0.18,0.31,0.31}
\definecolor{DarkTurquoise}{rgb}{0.00,0.81,0.82}
\definecolor{DarkViolet}{rgb}{0.58,0.00,0.83}
\definecolor{DeepPink1}{rgb}{1.00,0.08,0.58}
\definecolor{DeepPink2}{rgb}{0.93,0.07,0.54}
\definecolor{DeepPink3}{rgb}{0.80,0.06,0.46}
\definecolor{DeepPink4}{rgb}{0.55,0.04,0.31}
\definecolor{DeepPink}{rgb}{1.00,0.08,0.58}
\definecolor{DeepSkyBlue1}{rgb}{0.00,0.75,1.00}
\definecolor{DeepSkyBlue2}{rgb}{0.00,0.70,0.93}
\definecolor{DeepSkyBlue3}{rgb}{0.00,0.60,0.80}
\definecolor{DeepSkyBlue4}{rgb}{0.00,0.41,0.55}
\definecolor{DeepSkyBlue}{rgb}{0.00,0.75,1.00}
\definecolor{DimGray}{rgb}{0.41,0.41,0.41}
\definecolor{DimGrey}{rgb}{0.41,0.41,0.41}
\definecolor{DodgerBlue1}{rgb}{0.12,0.56,1.00}
\definecolor{DodgerBlue2}{rgb}{0.11,0.53,0.93}
\definecolor{DodgerBlue3}{rgb}{0.09,0.45,0.80}
\definecolor{DodgerBlue4}{rgb}{0.06,0.31,0.55}
\definecolor{DodgerBlue}{rgb}{0.12,0.56,1.00}
\definecolor{FloralWhite}{rgb}{1.00,0.98,0.94}
\definecolor{ForestGreen}{rgb}{0.13,0.55,0.13}
\definecolor{GhostWhite}{rgb}{0.97,0.97,1.00}
\definecolor{GreenYellow}{rgb}{0.68,1.00,0.18}
\definecolor{HotPink1}{rgb}{1.00,0.43,0.71}
\definecolor{HotPink2}{rgb}{0.93,0.42,0.65}
\definecolor{HotPink3}{rgb}{0.80,0.38,0.56}
\definecolor{HotPink4}{rgb}{0.55,0.23,0.38}
\definecolor{HotPink}{rgb}{1.00,0.41,0.71}
\definecolor{IndianRed1}{rgb}{1.00,0.42,0.42}
\definecolor{IndianRed2}{rgb}{0.93,0.39,0.39}
\definecolor{IndianRed3}{rgb}{0.80,0.33,0.33}
\definecolor{IndianRed4}{rgb}{0.55,0.23,0.23}
\definecolor{IndianRed}{rgb}{0.80,0.36,0.36}
\definecolor{LavenderBlush1}{rgb}{1.00,0.94,0.96}
\definecolor{LavenderBlush2}{rgb}{0.93,0.88,0.90}
\definecolor{LavenderBlush3}{rgb}{0.80,0.76,0.77}
\definecolor{LavenderBlush4}{rgb}{0.55,0.51,0.53}
\definecolor{LavenderBlush}{rgb}{1.00,0.94,0.96}
\definecolor{LawnGreen}{rgb}{0.49,0.99,0.00}
\definecolor{LemonChiffon1}{rgb}{1.00,0.98,0.80}
\definecolor{LemonChiffon2}{rgb}{0.93,0.91,0.75}
\definecolor{LemonChiffon3}{rgb}{0.80,0.79,0.65}
\definecolor{LemonChiffon4}{rgb}{0.55,0.54,0.44}
\definecolor{LemonChiffon}{rgb}{1.00,0.98,0.80}
\definecolor{LightBlue1}{rgb}{0.75,0.94,1.00}
\definecolor{LightBlue2}{rgb}{0.70,0.87,0.93}
\definecolor{LightBlue3}{rgb}{0.60,0.75,0.80}
\definecolor{LightBlue4}{rgb}{0.41,0.51,0.55}
\definecolor{LightBlue}{rgb}{0.68,0.85,0.90}
\definecolor{LightCoral}{rgb}{0.94,0.50,0.50}
\definecolor{LightCyan1}{rgb}{0.88,1.00,1.00}
\definecolor{LightCyan2}{rgb}{0.82,0.93,0.93}
\definecolor{LightCyan3}{rgb}{0.71,0.80,0.80}
\definecolor{LightCyan4}{rgb}{0.48,0.55,0.55}
\definecolor{LightCyan}{rgb}{0.88,1.00,1.00}
\definecolor{LightGoldenrod1}{rgb}{1.00,0.93,0.55}
\definecolor{LightGoldenrod2}{rgb}{0.93,0.86,0.51}
\definecolor{LightGoldenrod3}{rgb}{0.80,0.75,0.44}
\definecolor{LightGoldenrod4}{rgb}{0.55,0.51,0.30}
\definecolor{LightGoldenrodYellow}{rgb}{0.98,0.98,0.82}
\definecolor{LightGoldenrod}{rgb}{0.93,0.87,0.51}
\definecolor{LightGray}{rgb}{0.83,0.83,0.83}
\definecolor{LightGreen}{rgb}{0.56,0.93,0.56}
\definecolor{LightGrey}{rgb}{0.83,0.83,0.83}
\definecolor{LightPink1}{rgb}{1.00,0.68,0.73}
\definecolor{LightPink2}{rgb}{0.93,0.64,0.68}
\definecolor{LightPink3}{rgb}{0.80,0.55,0.58}
\definecolor{LightPink4}{rgb}{0.55,0.37,0.40}
\definecolor{LightPink}{rgb}{1.00,0.71,0.76}
\definecolor{LightSalmon1}{rgb}{1.00,0.63,0.48}
\definecolor{LightSalmon2}{rgb}{0.93,0.58,0.45}
\definecolor{LightSalmon3}{rgb}{0.80,0.51,0.38}
\definecolor{LightSalmon4}{rgb}{0.55,0.34,0.26}
\definecolor{LightSalmon}{rgb}{1.00,0.63,0.48}
\definecolor{LightSeaGreen}{rgb}{0.13,0.70,0.67}
\definecolor{LightSkyBlue1}{rgb}{0.69,0.89,1.00}
\definecolor{LightSkyBlue2}{rgb}{0.64,0.83,0.93}
\definecolor{LightSkyBlue3}{rgb}{0.55,0.71,0.80}
\definecolor{LightSkyBlue4}{rgb}{0.38,0.48,0.55}
\definecolor{LightSkyBlue}{rgb}{0.53,0.81,0.98}
\definecolor{LightSlateBlue}{rgb}{0.52,0.44,1.00}
\definecolor{LightSlateGray}{rgb}{0.47,0.53,0.60}
\definecolor{LightSlateGrey}{rgb}{0.47,0.53,0.60}
\definecolor{LightSteelBlue1}{rgb}{0.79,0.88,1.00}
\definecolor{LightSteelBlue2}{rgb}{0.74,0.82,0.93}
\definecolor{LightSteelBlue3}{rgb}{0.64,0.71,0.80}
\definecolor{LightSteelBlue4}{rgb}{0.43,0.48,0.55}
\definecolor{LightSteelBlue}{rgb}{0.69,0.77,0.87}
\definecolor{LightYellow1}{rgb}{1.00,1.00,0.88}
\definecolor{LightYellow2}{rgb}{0.93,0.93,0.82}
\definecolor{LightYellow3}{rgb}{0.80,0.80,0.71}
\definecolor{LightYellow4}{rgb}{0.55,0.55,0.48}
\definecolor{LightYellow}{rgb}{1.00,1.00,0.88}
\definecolor{LimeGreen}{rgb}{0.20,0.80,0.20}
\definecolor{MediumAquamarine}{rgb}{0.40,0.80,0.67}
\definecolor{MediumBlue}{rgb}{0.00,0.00,0.80}
\definecolor{MediumOrchid1}{rgb}{0.88,0.40,1.00}
\definecolor{MediumOrchid2}{rgb}{0.82,0.37,0.93}
\definecolor{MediumOrchid3}{rgb}{0.71,0.32,0.80}
\definecolor{MediumOrchid4}{rgb}{0.48,0.22,0.55}
\definecolor{MediumOrchid}{rgb}{0.73,0.33,0.83}
\definecolor{MediumPurple1}{rgb}{0.67,0.51,1.00}
\definecolor{MediumPurple2}{rgb}{0.62,0.47,0.93}
\definecolor{MediumPurple3}{rgb}{0.54,0.41,0.80}
\definecolor{MediumPurple4}{rgb}{0.36,0.28,0.55}
\definecolor{MediumPurple}{rgb}{0.58,0.44,0.86}
\definecolor{MediumSeaGreen}{rgb}{0.24,0.70,0.44}
\definecolor{MediumSlateBlue}{rgb}{0.48,0.41,0.93}
\definecolor{MediumSpringGreen}{rgb}{0.00,0.98,0.60}
\definecolor{MediumTurquoise}{rgb}{0.28,0.82,0.80}
\definecolor{MediumVioletRed}{rgb}{0.78,0.08,0.52}
\definecolor{MidnightBlue}{rgb}{0.10,0.10,0.44}
\definecolor{MintCream}{rgb}{0.96,1.00,0.98}
\definecolor{MistyRose1}{rgb}{1.00,0.89,0.88}
\definecolor{MistyRose2}{rgb}{0.93,0.84,0.82}
\definecolor{MistyRose3}{rgb}{0.80,0.72,0.71}
\definecolor{MistyRose4}{rgb}{0.55,0.49,0.48}
\definecolor{MistyRose}{rgb}{1.00,0.89,0.88}
\definecolor{NavajoWhite1}{rgb}{1.00,0.87,0.68}
\definecolor{NavajoWhite2}{rgb}{0.93,0.81,0.63}
\definecolor{NavajoWhite3}{rgb}{0.80,0.70,0.55}
\definecolor{NavajoWhite4}{rgb}{0.55,0.47,0.37}
\definecolor{NavajoWhite}{rgb}{1.00,0.87,0.68}
\definecolor{NavyBlue}{rgb}{0.00,0.00,0.50}
\definecolor{OldLace}{rgb}{0.99,0.96,0.90}
\definecolor{OliveDrab1}{rgb}{0.75,1.00,0.24}
\definecolor{OliveDrab2}{rgb}{0.70,0.93,0.23}
\definecolor{OliveDrab3}{rgb}{0.60,0.80,0.20}
\definecolor{OliveDrab4}{rgb}{0.41,0.55,0.13}
\definecolor{OliveDrab}{rgb}{0.42,0.56,0.14}
\definecolor{OrangeRed1}{rgb}{1.00,0.27,0.00}
\definecolor{OrangeRed2}{rgb}{0.93,0.25,0.00}
\definecolor{OrangeRed3}{rgb}{0.80,0.22,0.00}
\definecolor{OrangeRed4}{rgb}{0.55,0.15,0.00}
\definecolor{OrangeRed}{rgb}{1.00,0.27,0.00}
\definecolor{PaleGoldenrod}{rgb}{0.93,0.91,0.67}
\definecolor{PaleGreen1}{rgb}{0.60,1.00,0.60}
\definecolor{PaleGreen2}{rgb}{0.56,0.93,0.56}
\definecolor{PaleGreen3}{rgb}{0.49,0.80,0.49}
\definecolor{PaleGreen4}{rgb}{0.33,0.55,0.33}
\definecolor{PaleGreen}{rgb}{0.60,0.98,0.60}
\definecolor{PaleTurquoise1}{rgb}{0.73,1.00,1.00}
\definecolor{PaleTurquoise2}{rgb}{0.68,0.93,0.93}
\definecolor{PaleTurquoise3}{rgb}{0.59,0.80,0.80}
\definecolor{PaleTurquoise4}{rgb}{0.40,0.55,0.55}
\definecolor{PaleTurquoise}{rgb}{0.69,0.93,0.93}
\definecolor{PaleVioletRed1}{rgb}{1.00,0.51,0.67}
\definecolor{PaleVioletRed2}{rgb}{0.93,0.47,0.62}
\definecolor{PaleVioletRed3}{rgb}{0.80,0.41,0.54}
\definecolor{PaleVioletRed4}{rgb}{0.55,0.28,0.36}
\definecolor{PaleVioletRed}{rgb}{0.86,0.44,0.58}
\definecolor{PapayaWhip}{rgb}{1.00,0.94,0.84}
\definecolor{PeachPuff1}{rgb}{1.00,0.85,0.73}
\definecolor{PeachPuff2}{rgb}{0.93,0.80,0.68}
\definecolor{PeachPuff3}{rgb}{0.80,0.69,0.58}
\definecolor{PeachPuff4}{rgb}{0.55,0.47,0.40}
\definecolor{PeachPuff}{rgb}{1.00,0.85,0.73}
\definecolor{PowderBlue}{rgb}{0.69,0.88,0.90}
\definecolor{RosyBrown1}{rgb}{1.00,0.76,0.76}
\definecolor{RosyBrown2}{rgb}{0.93,0.71,0.71}
\definecolor{RosyBrown3}{rgb}{0.80,0.61,0.61}
\definecolor{RosyBrown4}{rgb}{0.55,0.41,0.41}
\definecolor{RosyBrown}{rgb}{0.74,0.56,0.56}
\definecolor{RoyalBlue1}{rgb}{0.28,0.46,1.00}
\definecolor{RoyalBlue2}{rgb}{0.26,0.43,0.93}
\definecolor{RoyalBlue3}{rgb}{0.23,0.37,0.80}
\definecolor{RoyalBlue4}{rgb}{0.15,0.25,0.55}
\definecolor{RoyalBlue}{rgb}{0.25,0.41,0.88}
\definecolor{SaddleBrown}{rgb}{0.55,0.27,0.07}
\definecolor{SandyBrown}{rgb}{0.96,0.64,0.38}
\definecolor{SeaGreen1}{rgb}{0.33,1.00,0.62}
\definecolor{SeaGreen2}{rgb}{0.31,0.93,0.58}
\definecolor{SeaGreen3}{rgb}{0.26,0.80,0.50}
\definecolor{SeaGreen4}{rgb}{0.18,0.55,0.34}
\definecolor{SeaGreen}{rgb}{0.18,0.55,0.34}
\definecolor{SkyBlue1}{rgb}{0.53,0.81,1.00}
\definecolor{SkyBlue2}{rgb}{0.49,0.75,0.93}
\definecolor{SkyBlue3}{rgb}{0.42,0.65,0.80}
\definecolor{SkyBlue4}{rgb}{0.29,0.44,0.55}
\definecolor{SkyBlue}{rgb}{0.53,0.81,0.92}
\definecolor{SlateBlue1}{rgb}{0.51,0.44,1.00}
\definecolor{SlateBlue2}{rgb}{0.48,0.40,0.93}
\definecolor{SlateBlue3}{rgb}{0.41,0.35,0.80}
\definecolor{SlateBlue4}{rgb}{0.28,0.24,0.55}
\definecolor{SlateBlue}{rgb}{0.42,0.35,0.80}
\definecolor{SlateGray1}{rgb}{0.78,0.89,1.00}
\definecolor{SlateGray2}{rgb}{0.73,0.83,0.93}
\definecolor{SlateGray3}{rgb}{0.62,0.71,0.80}
\definecolor{SlateGray4}{rgb}{0.42,0.48,0.55}
\definecolor{SlateGray}{rgb}{0.44,0.50,0.56}
\definecolor{SlateGrey}{rgb}{0.44,0.50,0.56}
\definecolor{SpringGreen1}{rgb}{0.00,1.00,0.50}
\definecolor{SpringGreen2}{rgb}{0.00,0.93,0.46}
\definecolor{SpringGreen3}{rgb}{0.00,0.80,0.40}
\definecolor{SpringGreen4}{rgb}{0.00,0.55,0.27}
\definecolor{SpringGreen}{rgb}{0.00,1.00,0.50}
\definecolor{SteelBlue1}{rgb}{0.39,0.72,1.00}
\definecolor{SteelBlue2}{rgb}{0.36,0.67,0.93}
\definecolor{SteelBlue3}{rgb}{0.31,0.58,0.80}
\definecolor{SteelBlue4}{rgb}{0.21,0.39,0.55}
\definecolor{SteelBlue}{rgb}{0.27,0.51,0.71}
\definecolor{VioletRed1}{rgb}{1.00,0.24,0.59}
\definecolor{VioletRed2}{rgb}{0.93,0.23,0.55}
\definecolor{VioletRed3}{rgb}{0.80,0.20,0.47}
\definecolor{VioletRed4}{rgb}{0.55,0.13,0.32}
\definecolor{VioletRed}{rgb}{0.82,0.13,0.56}
\definecolor{WhiteSmoke}{rgb}{0.96,0.96,0.96}
\definecolor{YellowGreen}{rgb}{0.60,0.80,0.20}
\definecolor{aliceblue}{rgb}{0.94,0.97,1.00}
\definecolor{antiquewhite}{rgb}{0.98,0.92,0.84}
\definecolor{aquamarine1}{rgb}{0.50,1.00,0.83}
\definecolor{aquamarine2}{rgb}{0.46,0.93,0.78}
\definecolor{aquamarine3}{rgb}{0.40,0.80,0.67}
\definecolor{aquamarine4}{rgb}{0.27,0.55,0.45}
\definecolor{aquamarine}{rgb}{0.50,1.00,0.83}
\definecolor{azure1}{rgb}{0.94,1.00,1.00}
\definecolor{azure2}{rgb}{0.88,0.93,0.93}
\definecolor{azure3}{rgb}{0.76,0.80,0.80}
\definecolor{azure4}{rgb}{0.51,0.55,0.55}
\definecolor{azure}{rgb}{0.94,1.00,1.00}
\definecolor{beige}{rgb}{0.96,0.96,0.86}
\definecolor{bisque1}{rgb}{1.00,0.89,0.77}
\definecolor{bisque2}{rgb}{0.93,0.84,0.72}
\definecolor{bisque3}{rgb}{0.80,0.72,0.62}
\definecolor{bisque4}{rgb}{0.55,0.49,0.42}
\definecolor{bisque}{rgb}{1.00,0.89,0.77}
\definecolor{black}{rgb}{0.00,0.00,0.00}
\definecolor{blanchedalmond}{rgb}{1.00,0.92,0.80}
\definecolor{blue1}{rgb}{0.00,0.00,1.00}
\definecolor{blue2}{rgb}{0.00,0.00,0.93}
\definecolor{blue3}{rgb}{0.00,0.00,0.80}
\definecolor{blue4}{rgb}{0.00,0.00,0.55}
\definecolor{blueviolet}{rgb}{0.54,0.17,0.89}
\definecolor{blue}{rgb}{0.00,0.00,1.00}
\definecolor{brown1}{rgb}{1.00,0.25,0.25}
\definecolor{brown2}{rgb}{0.93,0.23,0.23}
\definecolor{brown3}{rgb}{0.80,0.20,0.20}
\definecolor{brown4}{rgb}{0.55,0.14,0.14}
\definecolor{brown}{rgb}{0.65,0.16,0.16}
\definecolor{burlywood1}{rgb}{1.00,0.83,0.61}
\definecolor{burlywood2}{rgb}{0.93,0.77,0.57}
\definecolor{burlywood3}{rgb}{0.80,0.67,0.49}
\definecolor{burlywood4}{rgb}{0.55,0.45,0.33}
\definecolor{burlywood}{rgb}{0.87,0.72,0.53}
\definecolor{cadetblue}{rgb}{0.37,0.62,0.63}
\definecolor{chartreuse1}{rgb}{0.50,1.00,0.00}
\definecolor{chartreuse2}{rgb}{0.46,0.93,0.00}
\definecolor{chartreuse3}{rgb}{0.40,0.80,0.00}
\definecolor{chartreuse4}{rgb}{0.27,0.55,0.00}
\definecolor{chartreuse}{rgb}{0.50,1.00,0.00}
\definecolor{chocolate1}{rgb}{1.00,0.50,0.14}
\definecolor{chocolate2}{rgb}{0.93,0.46,0.13}
\definecolor{chocolate3}{rgb}{0.80,0.40,0.11}
\definecolor{chocolate4}{rgb}{0.55,0.27,0.07}
\definecolor{chocolate}{rgb}{0.82,0.41,0.12}
\definecolor{coral1}{rgb}{1.00,0.45,0.34}
\definecolor{coral2}{rgb}{0.93,0.42,0.31}
\definecolor{coral3}{rgb}{0.80,0.36,0.27}
\definecolor{coral4}{rgb}{0.55,0.24,0.18}
\definecolor{coral}{rgb}{1.00,0.50,0.31}
\definecolor{cornflowerblue}{rgb}{0.39,0.58,0.93}
\definecolor{cornsilk1}{rgb}{1.00,0.97,0.86}
\definecolor{cornsilk2}{rgb}{0.93,0.91,0.80}
\definecolor{cornsilk3}{rgb}{0.80,0.78,0.69}
\definecolor{cornsilk4}{rgb}{0.55,0.53,0.47}
\definecolor{cornsilk}{rgb}{1.00,0.97,0.86}
\definecolor{cyan1}{rgb}{0.00,1.00,1.00}
\definecolor{cyan2}{rgb}{0.00,0.93,0.93}
\definecolor{cyan3}{rgb}{0.00,0.80,0.80}
\definecolor{cyan4}{rgb}{0.00,0.55,0.55}
\definecolor{cyan}{rgb}{0.00,1.00,1.00}
\definecolor{darkblue}{rgb}{0.00,0.00,0.55}
\definecolor{darkcyan}{rgb}{0.00,0.55,0.55}
\definecolor{darkgoldenrod}{rgb}{0.72,0.53,0.04}
\definecolor{darkgray}{rgb}{0.66,0.66,0.66}
\definecolor{darkgreen}{rgb}{0.00,0.39,0.00}
\definecolor{darkgrey}{rgb}{0.66,0.66,0.66}
\definecolor{darkkhaki}{rgb}{0.74,0.72,0.42}
\definecolor{darkmagenta}{rgb}{0.55,0.00,0.55}
\definecolor{darkolive}{rgb}{0.33,0.42,0.18}
\definecolor{darkorange}{rgb}{1.00,0.55,0.00}
\definecolor{darkorchid}{rgb}{0.60,0.20,0.80}
\definecolor{darkred}{rgb}{0.55,0.00,0.00}
\definecolor{darksalmon}{rgb}{0.91,0.59,0.48}
\definecolor{darksea}{rgb}{0.56,0.74,0.56}
\definecolor{darkslate}{rgb}{0.18,0.31,0.31}
\definecolor{darkslate}{rgb}{0.18,0.31,0.31}
\definecolor{darkslate}{rgb}{0.28,0.24,0.55}
\definecolor{darkturquoise}{rgb}{0.00,0.81,0.82}
\definecolor{darkviolet}{rgb}{0.58,0.00,0.83}
\definecolor{deeppink}{rgb}{1.00,0.08,0.58}
\definecolor{deepsky}{rgb}{0.00,0.75,1.00}
\definecolor{dimgray}{rgb}{0.41,0.41,0.41}
\definecolor{dimgrey}{rgb}{0.41,0.41,0.41}
\definecolor{dodgerblue}{rgb}{0.12,0.56,1.00}
\definecolor{firebrick1}{rgb}{1.00,0.19,0.19}
\definecolor{firebrick2}{rgb}{0.93,0.17,0.17}
\definecolor{firebrick3}{rgb}{0.80,0.15,0.15}
\definecolor{firebrick4}{rgb}{0.55,0.10,0.10}
\definecolor{firebrick}{rgb}{0.70,0.13,0.13}
\definecolor{floralwhite}{rgb}{1.00,0.98,0.94}
\definecolor{forestgreen}{rgb}{0.13,0.55,0.13}
\definecolor{gainsboro}{rgb}{0.86,0.86,0.86}
\definecolor{ghostwhite}{rgb}{0.97,0.97,1.00}
\definecolor{gold1}{rgb}{1.00,0.84,0.00}
\definecolor{gold2}{rgb}{0.93,0.79,0.00}
\definecolor{gold3}{rgb}{0.80,0.68,0.00}
\definecolor{gold4}{rgb}{0.55,0.46,0.00}
\definecolor{goldenrod1}{rgb}{1.00,0.76,0.15}
\definecolor{goldenrod2}{rgb}{0.93,0.71,0.13}
\definecolor{goldenrod3}{rgb}{0.80,0.61,0.11}
\definecolor{goldenrod4}{rgb}{0.55,0.41,0.08}
\definecolor{goldenrod}{rgb}{0.85,0.65,0.13}
\definecolor{gold}{rgb}{1.00,0.84,0.00}
\definecolor{gray0}{rgb}{0.00,0.00,0.00}
\definecolor{gray100}{rgb}{1.00,1.00,1.00}
\definecolor{gray10}{rgb}{0.10,0.10,0.10}
\definecolor{gray11}{rgb}{0.11,0.11,0.11}
\definecolor{gray12}{rgb}{0.12,0.12,0.12}
\definecolor{gray13}{rgb}{0.13,0.13,0.13}
\definecolor{gray14}{rgb}{0.14,0.14,0.14}
\definecolor{gray15}{rgb}{0.15,0.15,0.15}
\definecolor{gray16}{rgb}{0.16,0.16,0.16}
\definecolor{gray17}{rgb}{0.17,0.17,0.17}
\definecolor{gray18}{rgb}{0.18,0.18,0.18}
\definecolor{gray19}{rgb}{0.19,0.19,0.19}
\definecolor{gray1}{rgb}{0.01,0.01,0.01}
\definecolor{gray20}{rgb}{0.20,0.20,0.20}
\definecolor{gray21}{rgb}{0.21,0.21,0.21}
\definecolor{gray22}{rgb}{0.22,0.22,0.22}
\definecolor{gray23}{rgb}{0.23,0.23,0.23}
\definecolor{gray24}{rgb}{0.24,0.24,0.24}
\definecolor{gray25}{rgb}{0.25,0.25,0.25}
\definecolor{gray26}{rgb}{0.26,0.26,0.26}
\definecolor{gray27}{rgb}{0.27,0.27,0.27}
\definecolor{gray28}{rgb}{0.28,0.28,0.28}
\definecolor{gray29}{rgb}{0.29,0.29,0.29}
\definecolor{gray2}{rgb}{0.02,0.02,0.02}
\definecolor{gray30}{rgb}{0.30,0.30,0.30}
\definecolor{gray31}{rgb}{0.31,0.31,0.31}
\definecolor{gray32}{rgb}{0.32,0.32,0.32}
\definecolor{gray33}{rgb}{0.33,0.33,0.33}
\definecolor{gray34}{rgb}{0.34,0.34,0.34}
\definecolor{gray35}{rgb}{0.35,0.35,0.35}
\definecolor{gray36}{rgb}{0.36,0.36,0.36}
\definecolor{gray37}{rgb}{0.37,0.37,0.37}
\definecolor{gray38}{rgb}{0.38,0.38,0.38}
\definecolor{gray39}{rgb}{0.39,0.39,0.39}
\definecolor{gray3}{rgb}{0.03,0.03,0.03}
\definecolor{gray40}{rgb}{0.40,0.40,0.40}
\definecolor{gray41}{rgb}{0.41,0.41,0.41}
\definecolor{gray42}{rgb}{0.42,0.42,0.42}
\definecolor{gray43}{rgb}{0.43,0.43,0.43}
\definecolor{gray44}{rgb}{0.44,0.44,0.44}
\definecolor{gray45}{rgb}{0.45,0.45,0.45}
\definecolor{gray46}{rgb}{0.46,0.46,0.46}
\definecolor{gray47}{rgb}{0.47,0.47,0.47}
\definecolor{gray48}{rgb}{0.48,0.48,0.48}
\definecolor{gray49}{rgb}{0.49,0.49,0.49}
\definecolor{gray4}{rgb}{0.04,0.04,0.04}
\definecolor{gray50}{rgb}{0.50,0.50,0.50}
\definecolor{gray51}{rgb}{0.51,0.51,0.51}
\definecolor{gray52}{rgb}{0.52,0.52,0.52}
\definecolor{gray53}{rgb}{0.53,0.53,0.53}
\definecolor{gray54}{rgb}{0.54,0.54,0.54}
\definecolor{gray55}{rgb}{0.55,0.55,0.55}
\definecolor{gray56}{rgb}{0.56,0.56,0.56}
\definecolor{gray57}{rgb}{0.57,0.57,0.57}
\definecolor{gray58}{rgb}{0.58,0.58,0.58}
\definecolor{gray59}{rgb}{0.59,0.59,0.59}
\definecolor{gray5}{rgb}{0.05,0.05,0.05}
\definecolor{gray60}{rgb}{0.60,0.60,0.60}
\definecolor{gray61}{rgb}{0.61,0.61,0.61}
\definecolor{gray62}{rgb}{0.62,0.62,0.62}
\definecolor{gray63}{rgb}{0.63,0.63,0.63}
\definecolor{gray64}{rgb}{0.64,0.64,0.64}
\definecolor{gray65}{rgb}{0.65,0.65,0.65}
\definecolor{gray66}{rgb}{0.66,0.66,0.66}
\definecolor{gray67}{rgb}{0.67,0.67,0.67}
\definecolor{gray68}{rgb}{0.68,0.68,0.68}
\definecolor{gray69}{rgb}{0.69,0.69,0.69}
\definecolor{gray6}{rgb}{0.06,0.06,0.06}
\definecolor{gray70}{rgb}{0.70,0.70,0.70}
\definecolor{gray71}{rgb}{0.71,0.71,0.71}
\definecolor{gray72}{rgb}{0.72,0.72,0.72}
\definecolor{gray73}{rgb}{0.73,0.73,0.73}
\definecolor{gray74}{rgb}{0.74,0.74,0.74}
\definecolor{gray75}{rgb}{0.75,0.75,0.75}
\definecolor{gray76}{rgb}{0.76,0.76,0.76}
\definecolor{gray77}{rgb}{0.77,0.77,0.77}
\definecolor{gray78}{rgb}{0.78,0.78,0.78}
\definecolor{gray79}{rgb}{0.79,0.79,0.79}
\definecolor{gray7}{rgb}{0.07,0.07,0.07}
\definecolor{gray80}{rgb}{0.80,0.80,0.80}
\definecolor{gray81}{rgb}{0.81,0.81,0.81}
\definecolor{gray82}{rgb}{0.82,0.82,0.82}
\definecolor{gray83}{rgb}{0.83,0.83,0.83}
\definecolor{gray84}{rgb}{0.84,0.84,0.84}
\definecolor{gray85}{rgb}{0.85,0.85,0.85}
\definecolor{gray86}{rgb}{0.86,0.86,0.86}
\definecolor{gray87}{rgb}{0.87,0.87,0.87}
\definecolor{gray88}{rgb}{0.88,0.88,0.88}
\definecolor{gray89}{rgb}{0.89,0.89,0.89}
\definecolor{gray8}{rgb}{0.08,0.08,0.08}
\definecolor{gray90}{rgb}{0.90,0.90,0.90}
\definecolor{gray91}{rgb}{0.91,0.91,0.91}
\definecolor{gray92}{rgb}{0.92,0.92,0.92}
\definecolor{gray93}{rgb}{0.93,0.93,0.93}
\definecolor{gray94}{rgb}{0.94,0.94,0.94}
\definecolor{gray95}{rgb}{0.95,0.95,0.95}
\definecolor{gray96}{rgb}{0.96,0.96,0.96}
\definecolor{gray97}{rgb}{0.97,0.97,0.97}
\definecolor{gray98}{rgb}{0.98,0.98,0.98}
\definecolor{gray99}{rgb}{0.99,0.99,0.99}
\definecolor{gray9}{rgb}{0.09,0.09,0.09}
\definecolor{gray}{rgb}{0.75,0.75,0.75}
\definecolor{green1}{rgb}{0.00,1.00,0.00}
\definecolor{green2}{rgb}{0.00,0.93,0.00}
\definecolor{green3}{rgb}{0.00,0.80,0.00}
\definecolor{green4}{rgb}{0.00,0.55,0.00}
\definecolor{greenyellow}{rgb}{0.68,1.00,0.18}
\definecolor{green}{rgb}{0.00,1.00,0.00}
\definecolor{grey0}{rgb}{0.00,0.00,0.00}
\definecolor{grey100}{rgb}{1.00,1.00,1.00}
\definecolor{grey10}{rgb}{0.10,0.10,0.10}
\definecolor{grey11}{rgb}{0.11,0.11,0.11}
\definecolor{grey12}{rgb}{0.12,0.12,0.12}
\definecolor{grey13}{rgb}{0.13,0.13,0.13}
\definecolor{grey14}{rgb}{0.14,0.14,0.14}
\definecolor{grey15}{rgb}{0.15,0.15,0.15}
\definecolor{grey16}{rgb}{0.16,0.16,0.16}
\definecolor{grey17}{rgb}{0.17,0.17,0.17}
\definecolor{grey18}{rgb}{0.18,0.18,0.18}
\definecolor{grey19}{rgb}{0.19,0.19,0.19}
\definecolor{grey1}{rgb}{0.01,0.01,0.01}
\definecolor{grey20}{rgb}{0.20,0.20,0.20}
\definecolor{grey21}{rgb}{0.21,0.21,0.21}
\definecolor{grey22}{rgb}{0.22,0.22,0.22}
\definecolor{grey23}{rgb}{0.23,0.23,0.23}
\definecolor{grey24}{rgb}{0.24,0.24,0.24}
\definecolor{grey25}{rgb}{0.25,0.25,0.25}
\definecolor{grey26}{rgb}{0.26,0.26,0.26}
\definecolor{grey27}{rgb}{0.27,0.27,0.27}
\definecolor{grey28}{rgb}{0.28,0.28,0.28}
\definecolor{grey29}{rgb}{0.29,0.29,0.29}
\definecolor{grey2}{rgb}{0.02,0.02,0.02}
\definecolor{grey30}{rgb}{0.30,0.30,0.30}
\definecolor{grey31}{rgb}{0.31,0.31,0.31}
\definecolor{grey32}{rgb}{0.32,0.32,0.32}
\definecolor{grey33}{rgb}{0.33,0.33,0.33}
\definecolor{grey34}{rgb}{0.34,0.34,0.34}
\definecolor{grey35}{rgb}{0.35,0.35,0.35}
\definecolor{grey36}{rgb}{0.36,0.36,0.36}
\definecolor{grey37}{rgb}{0.37,0.37,0.37}
\definecolor{grey38}{rgb}{0.38,0.38,0.38}
\definecolor{grey39}{rgb}{0.39,0.39,0.39}
\definecolor{grey3}{rgb}{0.03,0.03,0.03}
\definecolor{grey40}{rgb}{0.40,0.40,0.40}
\definecolor{grey41}{rgb}{0.41,0.41,0.41}
\definecolor{grey42}{rgb}{0.42,0.42,0.42}
\definecolor{grey43}{rgb}{0.43,0.43,0.43}
\definecolor{grey44}{rgb}{0.44,0.44,0.44}
\definecolor{grey45}{rgb}{0.45,0.45,0.45}
\definecolor{grey46}{rgb}{0.46,0.46,0.46}
\definecolor{grey47}{rgb}{0.47,0.47,0.47}
\definecolor{grey48}{rgb}{0.48,0.48,0.48}
\definecolor{grey49}{rgb}{0.49,0.49,0.49}
\definecolor{grey4}{rgb}{0.04,0.04,0.04}
\definecolor{grey50}{rgb}{0.50,0.50,0.50}
\definecolor{grey51}{rgb}{0.51,0.51,0.51}
\definecolor{grey52}{rgb}{0.52,0.52,0.52}
\definecolor{grey53}{rgb}{0.53,0.53,0.53}
\definecolor{grey54}{rgb}{0.54,0.54,0.54}
\definecolor{grey55}{rgb}{0.55,0.55,0.55}
\definecolor{grey56}{rgb}{0.56,0.56,0.56}
\definecolor{grey57}{rgb}{0.57,0.57,0.57}
\definecolor{grey58}{rgb}{0.58,0.58,0.58}
\definecolor{grey59}{rgb}{0.59,0.59,0.59}
\definecolor{grey5}{rgb}{0.05,0.05,0.05}
\definecolor{grey60}{rgb}{0.60,0.60,0.60}
\definecolor{grey61}{rgb}{0.61,0.61,0.61}
\definecolor{grey62}{rgb}{0.62,0.62,0.62}
\definecolor{grey63}{rgb}{0.63,0.63,0.63}
\definecolor{grey64}{rgb}{0.64,0.64,0.64}
\definecolor{grey65}{rgb}{0.65,0.65,0.65}
\definecolor{grey66}{rgb}{0.66,0.66,0.66}
\definecolor{grey67}{rgb}{0.67,0.67,0.67}
\definecolor{grey68}{rgb}{0.68,0.68,0.68}
\definecolor{grey69}{rgb}{0.69,0.69,0.69}
\definecolor{grey6}{rgb}{0.06,0.06,0.06}
\definecolor{grey70}{rgb}{0.70,0.70,0.70}
\definecolor{grey71}{rgb}{0.71,0.71,0.71}
\definecolor{grey72}{rgb}{0.72,0.72,0.72}
\definecolor{grey73}{rgb}{0.73,0.73,0.73}
\definecolor{grey74}{rgb}{0.74,0.74,0.74}
\definecolor{grey75}{rgb}{0.75,0.75,0.75}
\definecolor{grey76}{rgb}{0.76,0.76,0.76}
\definecolor{grey77}{rgb}{0.77,0.77,0.77}
\definecolor{grey78}{rgb}{0.78,0.78,0.78}
\definecolor{grey79}{rgb}{0.79,0.79,0.79}
\definecolor{grey7}{rgb}{0.07,0.07,0.07}
\definecolor{grey80}{rgb}{0.80,0.80,0.80}
\definecolor{grey81}{rgb}{0.81,0.81,0.81}
\definecolor{grey82}{rgb}{0.82,0.82,0.82}
\definecolor{grey83}{rgb}{0.83,0.83,0.83}
\definecolor{grey84}{rgb}{0.84,0.84,0.84}
\definecolor{grey85}{rgb}{0.85,0.85,0.85}
\definecolor{grey86}{rgb}{0.86,0.86,0.86}
\definecolor{grey87}{rgb}{0.87,0.87,0.87}
\definecolor{grey88}{rgb}{0.88,0.88,0.88}
\definecolor{grey89}{rgb}{0.89,0.89,0.89}
\definecolor{grey8}{rgb}{0.08,0.08,0.08}
\definecolor{grey90}{rgb}{0.90,0.90,0.90}
\definecolor{grey91}{rgb}{0.91,0.91,0.91}
\definecolor{grey92}{rgb}{0.92,0.92,0.92}
\definecolor{grey93}{rgb}{0.93,0.93,0.93}
\definecolor{grey94}{rgb}{0.94,0.94,0.94}
\definecolor{grey95}{rgb}{0.95,0.95,0.95}
\definecolor{grey96}{rgb}{0.96,0.96,0.96}
\definecolor{grey97}{rgb}{0.97,0.97,0.97}
\definecolor{grey98}{rgb}{0.98,0.98,0.98}
\definecolor{grey99}{rgb}{0.99,0.99,0.99}
\definecolor{grey9}{rgb}{0.09,0.09,0.09}
\definecolor{grey}{rgb}{0.75,0.75,0.75}
\definecolor{honeydew1}{rgb}{0.94,1.00,0.94}
\definecolor{honeydew2}{rgb}{0.88,0.93,0.88}
\definecolor{honeydew3}{rgb}{0.76,0.80,0.76}
\definecolor{honeydew4}{rgb}{0.51,0.55,0.51}
\definecolor{honeydew}{rgb}{0.94,1.00,0.94}
\definecolor{hotpink}{rgb}{1.00,0.41,0.71}
\definecolor{indianred}{rgb}{0.80,0.36,0.36}
\definecolor{ivory1}{rgb}{1.00,1.00,0.94}
\definecolor{ivory2}{rgb}{0.93,0.93,0.88}
\definecolor{ivory3}{rgb}{0.80,0.80,0.76}
\definecolor{ivory4}{rgb}{0.55,0.55,0.51}
\definecolor{ivory}{rgb}{1.00,1.00,0.94}
\definecolor{khaki1}{rgb}{1.00,0.96,0.56}
\definecolor{khaki2}{rgb}{0.93,0.90,0.52}
\definecolor{khaki3}{rgb}{0.80,0.78,0.45}
\definecolor{khaki4}{rgb}{0.55,0.53,0.31}
\definecolor{khaki}{rgb}{0.94,0.90,0.55}
\definecolor{lavenderblush}{rgb}{1.00,0.94,0.96}
\definecolor{lavender}{rgb}{0.90,0.90,0.98}
\definecolor{lawngreen}{rgb}{0.49,0.99,0.00}
\definecolor{lemonchiffon}{rgb}{1.00,0.98,0.80}
\definecolor{lightblue}{rgb}{0.68,0.85,0.90}
\definecolor{lightcoral}{rgb}{0.94,0.50,0.50}
\definecolor{lightcyan}{rgb}{0.88,1.00,1.00}
\definecolor{lightgoldenrod}{rgb}{0.93,0.87,0.51}
\definecolor{lightgoldenrod}{rgb}{0.98,0.98,0.82}
\definecolor{lightgray}{rgb}{0.83,0.83,0.83}
\definecolor{lightgreen}{rgb}{0.56,0.93,0.56}
\definecolor{lightgrey}{rgb}{0.83,0.83,0.83}
\definecolor{lightpink}{rgb}{1.00,0.71,0.76}
\definecolor{lightsalmon}{rgb}{1.00,0.63,0.48}
\definecolor{lightsea}{rgb}{0.13,0.70,0.67}
\definecolor{lightsky}{rgb}{0.53,0.81,0.98}
\definecolor{lightslate}{rgb}{0.47,0.53,0.60}
\definecolor{lightslate}{rgb}{0.47,0.53,0.60}
\definecolor{lightslate}{rgb}{0.52,0.44,1.00}
\definecolor{lightsteel}{rgb}{0.69,0.77,0.87}
\definecolor{lightyellow}{rgb}{1.00,1.00,0.88}
\definecolor{limegreen}{rgb}{0.20,0.80,0.20}
\definecolor{linen}{rgb}{0.98,0.94,0.90}
\definecolor{magenta1}{rgb}{1.00,0.00,1.00}
\definecolor{magenta2}{rgb}{0.93,0.00,0.93}
\definecolor{magenta3}{rgb}{0.80,0.00,0.80}
\definecolor{magenta4}{rgb}{0.55,0.00,0.55}
\definecolor{magenta}{rgb}{1.00,0.00,1.00}
\definecolor{maroon1}{rgb}{1.00,0.20,0.70}
\definecolor{maroon2}{rgb}{0.93,0.19,0.65}
\definecolor{maroon3}{rgb}{0.80,0.16,0.56}
\definecolor{maroon4}{rgb}{0.55,0.11,0.38}
\definecolor{maroon}{rgb}{0.69,0.19,0.38}
\definecolor{mediumaquamarine}{rgb}{0.40,0.80,0.67}
\definecolor{mediumblue}{rgb}{0.00,0.00,0.80}
\definecolor{mediumorchid}{rgb}{0.73,0.33,0.83}
\definecolor{mediumpurple}{rgb}{0.58,0.44,0.86}
\definecolor{mediumsea}{rgb}{0.24,0.70,0.44}
\definecolor{mediumslate}{rgb}{0.48,0.41,0.93}
\definecolor{mediumspring}{rgb}{0.00,0.98,0.60}
\definecolor{mediumturquoise}{rgb}{0.28,0.82,0.80}
\definecolor{mediumviolet}{rgb}{0.78,0.08,0.52}
\definecolor{midnightblue}{rgb}{0.10,0.10,0.44}
\definecolor{mintcream}{rgb}{0.96,1.00,0.98}
\definecolor{mistyrose}{rgb}{1.00,0.89,0.88}
\definecolor{moccasin}{rgb}{1.00,0.89,0.71}
\definecolor{navajowhite}{rgb}{1.00,0.87,0.68}
\definecolor{navyblue}{rgb}{0.00,0.00,0.50}
\definecolor{navy}{rgb}{0.00,0.00,0.50}
\definecolor{oldlace}{rgb}{0.99,0.96,0.90}
\definecolor{olivedrab}{rgb}{0.42,0.56,0.14}
\definecolor{orange1}{rgb}{1.00,0.65,0.00}
\definecolor{orange2}{rgb}{0.93,0.60,0.00}
\definecolor{orange3}{rgb}{0.80,0.52,0.00}
\definecolor{orange4}{rgb}{0.55,0.35,0.00}
\definecolor{orangered}{rgb}{1.00,0.27,0.00}
\definecolor{orange}{rgb}{1.00,0.65,0.00}
\definecolor{orchid1}{rgb}{1.00,0.51,0.98}
\definecolor{orchid2}{rgb}{0.93,0.48,0.91}
\definecolor{orchid3}{rgb}{0.80,0.41,0.79}
\definecolor{orchid4}{rgb}{0.55,0.28,0.54}
\definecolor{orchid}{rgb}{0.85,0.44,0.84}
\definecolor{palegoldenrod}{rgb}{0.93,0.91,0.67}
\definecolor{palegreen}{rgb}{0.60,0.98,0.60}
\definecolor{paleturquoise}{rgb}{0.69,0.93,0.93}
\definecolor{paleviolet}{rgb}{0.86,0.44,0.58}
\definecolor{papayawhip}{rgb}{1.00,0.94,0.84}
\definecolor{peachpuff}{rgb}{1.00,0.85,0.73}
\definecolor{peru}{rgb}{0.80,0.52,0.25}
\definecolor{pink1}{rgb}{1.00,0.71,0.77}
\definecolor{pink2}{rgb}{0.93,0.66,0.72}
\definecolor{pink3}{rgb}{0.80,0.57,0.62}
\definecolor{pink4}{rgb}{0.55,0.39,0.42}
\definecolor{pink}{rgb}{1.00,0.75,0.80}
\definecolor{plum1}{rgb}{1.00,0.73,1.00}
\definecolor{plum2}{rgb}{0.93,0.68,0.93}
\definecolor{plum3}{rgb}{0.80,0.59,0.80}
\definecolor{plum4}{rgb}{0.55,0.40,0.55}
\definecolor{plum}{rgb}{0.87,0.63,0.87}
\definecolor{powderblue}{rgb}{0.69,0.88,0.90}
\definecolor{purple1}{rgb}{0.61,0.19,1.00}
\definecolor{purple2}{rgb}{0.57,0.17,0.93}
\definecolor{purple3}{rgb}{0.49,0.15,0.80}
\definecolor{purple4}{rgb}{0.33,0.10,0.55}
\definecolor{purple}{rgb}{0.63,0.13,0.94}
\definecolor{red1}{rgb}{1.00,0.00,0.00}
\definecolor{red2}{rgb}{0.93,0.00,0.00}
\definecolor{red3}{rgb}{0.80,0.00,0.00}
\definecolor{red4}{rgb}{0.55,0.00,0.00}
\definecolor{red}{rgb}{1.00,0.00,0.00}
\definecolor{rosybrown}{rgb}{0.74,0.56,0.56}
\definecolor{royalblue}{rgb}{0.25,0.41,0.88}
\definecolor{saddlebrown}{rgb}{0.55,0.27,0.07}
\definecolor{salmon1}{rgb}{1.00,0.55,0.41}
\definecolor{salmon2}{rgb}{0.93,0.51,0.38}
\definecolor{salmon3}{rgb}{0.80,0.44,0.33}
\definecolor{salmon4}{rgb}{0.55,0.30,0.22}
\definecolor{salmon}{rgb}{0.98,0.50,0.45}
\definecolor{sandybrown}{rgb}{0.96,0.64,0.38}
\definecolor{seagreen}{rgb}{0.18,0.55,0.34}
\definecolor{seashell1}{rgb}{1.00,0.96,0.93}
\definecolor{seashell2}{rgb}{0.93,0.90,0.87}
\definecolor{seashell3}{rgb}{0.80,0.77,0.75}
\definecolor{seashell4}{rgb}{0.55,0.53,0.51}
\definecolor{seashell}{rgb}{1.00,0.96,0.93}
\definecolor{sienna1}{rgb}{1.00,0.51,0.28}
\definecolor{sienna2}{rgb}{0.93,0.47,0.26}
\definecolor{sienna3}{rgb}{0.80,0.41,0.22}
\definecolor{sienna4}{rgb}{0.55,0.28,0.15}
\definecolor{sienna}{rgb}{0.63,0.32,0.18}
\definecolor{skyblue}{rgb}{0.53,0.81,0.92}
\definecolor{slateblue}{rgb}{0.42,0.35,0.80}
\definecolor{slategray}{rgb}{0.44,0.50,0.56}
\definecolor{slategrey}{rgb}{0.44,0.50,0.56}
\definecolor{snow1}{rgb}{1.00,0.98,0.98}
\definecolor{snow2}{rgb}{0.93,0.91,0.91}
\definecolor{snow3}{rgb}{0.80,0.79,0.79}
\definecolor{snow4}{rgb}{0.55,0.54,0.54}
\definecolor{snow}{rgb}{1.00,0.98,0.98}
\definecolor{springgreen}{rgb}{0.00,1.00,0.50}
\definecolor{steelblue}{rgb}{0.27,0.51,0.71}
\definecolor{tan1}{rgb}{1.00,0.65,0.31}
\definecolor{tan2}{rgb}{0.93,0.60,0.29}
\definecolor{tan3}{rgb}{0.80,0.52,0.25}
\definecolor{tan4}{rgb}{0.55,0.35,0.17}
\definecolor{tan}{rgb}{0.82,0.71,0.55}
\definecolor{thistle1}{rgb}{1.00,0.88,1.00}
\definecolor{thistle2}{rgb}{0.93,0.82,0.93}
\definecolor{thistle3}{rgb}{0.80,0.71,0.80}
\definecolor{thistle4}{rgb}{0.55,0.48,0.55}
\definecolor{thistle}{rgb}{0.85,0.75,0.85}
\definecolor{tomato1}{rgb}{1.00,0.39,0.28}
\definecolor{tomato2}{rgb}{0.93,0.36,0.26}
\definecolor{tomato3}{rgb}{0.80,0.31,0.22}
\definecolor{tomato4}{rgb}{0.55,0.21,0.15}
\definecolor{tomato}{rgb}{1.00,0.39,0.28}
\definecolor{turquoise1}{rgb}{0.00,0.96,1.00}
\definecolor{turquoise2}{rgb}{0.00,0.90,0.93}
\definecolor{turquoise3}{rgb}{0.00,0.77,0.80}
\definecolor{turquoise4}{rgb}{0.00,0.53,0.55}
\definecolor{turquoise}{rgb}{0.25,0.88,0.82}
\definecolor{violetred}{rgb}{0.82,0.13,0.56}
\definecolor{violet}{rgb}{0.93,0.51,0.93}
\definecolor{wheat1}{rgb}{1.00,0.91,0.73}
\definecolor{wheat2}{rgb}{0.93,0.85,0.68}
\definecolor{wheat3}{rgb}{0.80,0.73,0.59}
\definecolor{wheat4}{rgb}{0.55,0.49,0.40}
\definecolor{wheat}{rgb}{0.96,0.87,0.70}
\definecolor{whitesmoke}{rgb}{0.96,0.96,0.96}
\definecolor{white}{rgb}{1.00,1.00,1.00}
\definecolor{yellow1}{rgb}{1.00,1.00,0.00}
\definecolor{yellow2}{rgb}{0.93,0.93,0.00}
\definecolor{yellow3}{rgb}{0.80,0.80,0.00}
\definecolor{yellow4}{rgb}{0.55,0.55,0.00}
\definecolor{yellowgreen}{rgb}{0.60,0.80,0.20}
\definecolor{yellow}{rgb}{1.00,1.00,0.00}